\documentclass[prb,superscriptaddress,amsfonts,amssymb,amsmath,floats,twocolumn,aps,footinbib,shownopacs]{revtex4}
\usepackage[pdftex]{graphicx}
\usepackage{subfigure}
\usepackage{dsfont}
\usepackage{lipsum}
\usepackage{dcolumn}
\usepackage{bm}
\usepackage{color}
\usepackage{physics}
\usepackage{multirow}
\usepackage[pdftex,colorlinks=true, linkcolor=blue,citecolor=blue,filecolor=blue]{hyperref}
\usepackage{comment}
\newcommand{\sgn}{\operatorname{sgn}}
\begin{document}
\title{Stochastic semiclassical theory for non-equilibrium electron-phonon coupled systems}

\author{Antonio Picano$^\ast$} 
\affiliation{Department of Physics, University of Erlangen-Nuremberg, Staudtstra{\ss}e 7, 91058 Erlangen, Germany}
\author{Francesco Grandi$^\ast$} 
\affiliation{Department of Physics, University of Erlangen-Nuremberg, Staudtstra{\ss}e 7, 91058 Erlangen, Germany}
\affiliation{Institute for Theory of Statistical Physics, RWTH Aachen University, 52056 Aachen, Germany}
\author{Philipp Werner} 
\affiliation{Department of Physics, University of Fribourg, 1700 Fribourg, Switzerland}
\author{Martin Eckstein}
\affiliation{Department of Physics, University of Erlangen-Nuremberg, Staudtstra{\ss}e 7, 91058 Erlangen, Germany}

\begin{abstract}
We discuss a semiclassical approach to solve the quantum impurity model within non-equilibrium dynamical mean-field theory for electron-lattice models. The effect of electronic fluctuations on the phonon is kept beyond Ehrenfest dynamics, leading to a stochastic phonon evolution with damping and noise terms that are self-consistently determined by the electronic correlation functions in the fluctuating phonon field. Together with a solution of the electronic model based on a non-perturbative quantum Boltzmann equation, the approach can be used to address the coupled dynamics of the electrons and the lattice during photo-induced phase transitions. Results for the Anderson-Holstein model are benchmarked against numerically exact quantum Monte Carlo data. We find good agreement for the phonon distribution function at temperatures comparable to the charge ordering temperature. The general formulation can be extended to models with electron-electron interactions or multi-orbital systems. 
\end{abstract} 
\maketitle 

\section{\label{sec:Introduction}Introduction}

\let\thefootnote\relax\footnotetext{*These authors equally contributed to this work}

The coupled non-equilibrium dynamics of the electrons and the lattice in solids leads to processes which can be orders of magnitude slower  than the intrinsic microscopic timescale of the electrons,  including coherent amplitude modes in charge density wave systems \cite{Huber2014,Perfetti2006}, or photo-induced phase transitions \cite{Zong2018, Kogar2019, Maklar2021, Wall2018, PerezSalinas2022, Beaud2014} and non-thermal transitions to disordered and metastable states \cite{Gerasimenko2019, Ichikawa2011, Stojchevska2014}. The large separation between the fast electronic dynamics and  slow lattice dynamics poses a considerable challenge for numerical simulations, even for minimal models such as the Holstein or Hubbard-Holstein model. Direct wave-function-based techniques like exact diagonalization and matrix product state algorithms \cite{Jeckelmann1999, Ejima2009, Jansen2020} for electron-phonon coupled systems must cope with the large bosonic Hilbert space. Efficient procedures exist for the dilute limit of a few polarons \cite{Vidmar2011, DeFilippis2012, Dorfner2015, Bonca1999}, but simulations at finite electron density \cite{Sous2021, Jansen2021} remain restricted to short times.  For systems with large spatial dimension,  non-equilibrium dynamical mean-field theory (DMFT) \cite{Aoki2014} becomes the reference method. DMFT maps a lattice model with local electron-phonon coupling to a single-site impurity model. In equilibrium, this model can be solved using quantum Monte Carlo (QMC) techniques \cite{Assaad2007, Werner2007}, but non-equilibrium simulations usually rely on perturbative weak-coupling expansions \cite{Murakami2015, Randi2017} or the strong coupling expansion around the atomic limit \cite{Eckstein2010, Werner2013}.  The latter also provides a good starting point for the regime of strong electron-phonon coupling, but different ways of integrating out the phonon within this formalism correspond to additional approximations when used within low  orders of the expansion (such as the non-crossing approximation, NCA) \cite{ Werner2013, Werner2015, Chen2016}. The unbiased approach, on the other hand, which includes the phonon exactly in the atomic  limit of the impurity model \cite{Grandi2021b}, faces a steep increase of the computational cost with the size of the phonon Hilbert space and is therefore again restricted  to short times.

In this work, we explore an alternative approach to solve the impurity model for DMFT-based electron-lattice models up to long times, based on a semiclassical approximation \cite{KamenevBook, Mitra2005, Picano2021_CDW} to the phonon dynamics. A  straightforward semiclassical approach would use a mean-field decoupling of the electrons and phonons, leading to classical equations of motion for the phonons with a force determined by electronic expectation values (Ehrenfest dynamics) \cite{Yonemitsu2009}. Beyond that, thermal and quantum fluctuations can be expected to have a profound  impact on the lattice dynamics. One way to incorporate fluctuations into a semiclassical evolution is through a weighted average over initial states for the classical variable \cite{Petrovic2022, Weber2022, Osterkorn2022, Bakshi2022}. Here, we aim to include the back-action of electronic fluctuations on the phonon also during the dynamics. This leads to a stochastic phonon evolution, with damping and noise obtained self-consistently from the electronic system driven by the fluctuating phonon field. With this, one can potentially address all the relevant stages of a photo-induced phase transition, including the recovery of the ordered phase and the thermalization of the phonons to their final equilibrium statistical distribution. The approach still requires a description of the electronic state at long times, but for well-separated electron and lattice timescales one can rely on a quasi steady approximation obtained from a non-perturbative quantum Boltzmann equation \cite{Picano2021}, or possibly even an adiabatic approximation. Nevertheless, in this work we only analyze the equilibrium properties of the model, so that we can benchmark the stochastic semiclassical theory against a numerically exact method.

While the semiclassical approach misses quantum processes like the tunneling between polaronic configurations \cite{Mitra2005}, it can be expected to provide a suitable description at sufficiently high temperature. The approach has been used to  study the melting and recovery of the charge density wave in the Holstein model on a Bethe lattice \cite{Picano2021_CDW}, where it revealed that the recovery of the ordered state proceeds through a disordered polaron phase, in which the local order parameter is not representative of the mean.  This scenario for the photo-induced dynamics  differs from the phenomenology obtained by time-dependent Ginzburg Landau theory, and implies a very slow recovery dynamics. The aim of the present work is to provide a derivation of the semiclassical approach which extends to models with electron-electron interactions and multi-orbital systems, and could therefore be used to describe, e.g., correlated electrons coupled to Jahn-Teller phonons,  or realistic models for coupled electron-lattice dynamics in materials like VO$_2$  \cite{Grandi2019}. Moreover, for the Anderson Holstein impurity model, we  benchmark the approach against the exact solution in equilibrium, and compare to approximate solutions based on the self-consistent Migdal approximation and the NCA approximation. 

The outline of this paper is as follows. In Sec.~\ref{sec:Semiclassical}, we derive the stochastic semiclassical equations of motion for the lattice displacement in a quantum impurity model with general linear coupling of the phonon displacement to an electronic operator. In Sec.~\ref{sec:holstein_model}, we introduce the Anderson Holstein model (Sec.~\ref{sec:model}), we summarize the corresponding QMC formalism which is used for the exact benchmarks (Sec.~\ref{sec:MonteCarlo}), as well as the solution in weak (Sec.~\ref{sec:MigdalApprox}) and strong coupling perturbation theory (Sec.~\ref{sec:NCA}). Sec.~\ref{sec:semicl_eq_holstein} contains the stochastic semiclassical equations of motion for the Anderson Holstein model. In Sec.~\ref{sec:Results}, we compare the equilibrium phonon distributions and electronic spectral functions  obtained with the various techniques. A summary and conclusion are presented in Sec.~\ref{sec:Conclusion}.

\section{\label{sec:Semiclassical} Theory for the semiclassical equations of motion for the lattice displacement}

\subsection{Impurity action}
In this section, we provide a derivation of the stochastic semiclassical equations of motion for the local lattice distortion, starting from a generic impurity problem in which electrons interact with a local vibrational mode. The electronic part of the impurity Hamiltonian includes the hybridization with an electron reservoir, and possibly a local interaction (such as a Hubbard interaction). The precise form will be specified in the numerical examples below. The phonon is described in terms of the Hamiltonian
\begin{align} 
	H_{x}=\frac{1}{2}\Big(\Omega^2\hat X^2+\hat P^2\Big),
	\label{Hx}
\end{align}
with canonically conjugate variables $\hat X$ (displacement) and $\hat P$ (momentum).
The vibrational mode interacts with the electrons on the impurity with the linear  term 
\begin{align} 
	& H_{cx} = \sqrt{2\Omega} g \hat X \hat O,
	\label{Hcx}
\end{align}
where $\hat O$ is a generic local electronic operator. For example, in the Holstein model, $\hat O$ is the local electron density $\hat O=\sum_{\sigma} \hat c_{\sigma}^\dagger \hat  c_{\sigma}$, while  interesting settings in multi-orbital impurity models would include a displacement $\hat X$ which modifies the crystal field splitting (i.e., $\hat O$ is the occupation difference $\hat n_1- \hat n_2$ between orbitals), or the hybridization between two impurity orbitals ($\hat O=\hat c_1^\dagger \hat c_2+\hat c_{2}^\dagger \hat c_1$).

We aim to solve the problem using a path integral formulation on the Keldysh contour $\mathcal{C} = \mathcal{C}^+ \cup \mathcal{C}^-$, with $\mathcal{C}^+ = ( - t_{\rm max},  t_{\rm max})$ and $\mathcal{C}^- = (  t_{\rm max}, - t_{\rm max})$, $t_{\rm max}\to\infty$.  We will henceforth use a notation such that the contour time $\tau=t^{+}$ ($\tau=t^{-}$) denotes the physical time $t$ on the upper (lower) branch of $\mathcal{C}$. In order to derive stochastic equations of motion for the phonon variable,  we keep the discrete-time notation for the  path integral, and divide $\mathcal{C}$  into $2N-2$ time intervals of length $\delta_t$. The physical time $t$ takes $N$ equidistant values $t_1=-t_{\rm max}, ... , t_N=t_{\rm max}$, while the contour times $\tau_j$ run from $j=1,...,2N$, with  $\tau_1=-t_{\rm max}^+$, $\tau_N=t_{\rm max}^+$, $\tau_{N+1}=t_{\rm max}^-$, and $\tau_{2N}=-t_{\rm max}^-$. The action is given in terms of the displacement field $X_j=X(\tau_j)$, and Grassmann fields $c_{j}=c(\tau_j)$ $\bar c_{j}=\bar c(\tau_j)$ for the electrons. In addition to the time index $j$, electron operators carry orbital and/or spin indices, which are not shown for simplicity. The action is given by 
\begin{align}
	&S[\bar c,c,X ]=S_{cc}[\bar c,c] + S_{x}[X] + S_{cx}[\bar c,c,X], 
	\\
	\label{action_x}
	& S_{x} = \sum_{j = 1}^{2N-1} \frac{h_{j+1}}{2} \Big[ \Big( \frac{X_{j+1} - X_j}{h_{j+1}} \Big)^2 - \Omega^2 X_j^2  \Big], \\ 
	\label{action_cx1}
	& S_{cx} =  -\sqrt{2\Omega}g \sum_{j = 1}^{2N-1} h_{j+1} \,O_{j+1} X_{j},
\end{align}
where the term $S_{cc}[\bar c,c]$ is purely electronic, while $S_{x}$ and $S_{cx}$ incorporate the bare phonon and the electron phonon interaction, respectively; $h_j=\tau_{j}-\tau_{j-1}$ is the time step (which is $\pm\delta_t$ for $j$ on the upper/lower contour, and $h_{N+1}=0$). Moreover, $O_j$ is the representation of the operator $\hat O$ in terms of Grassmann variables, i.e., if $\hat O[\hat c^\dagger, \hat c]$ is a normal-ordered function of $\hat c^\dagger$ and $ \hat c$, then $O_{j}=O[\bar c_j,c_{j-1}]$. 

With the action, the partition function can be written as
\begin{align}
	Z=\int \mathcal{D}[X] \int\mathcal{D}[\bar c, c]\, e^{i S[\bar c,c,X ]},
\end{align}
from which time-dependent expectation values are obtained by taking derivatives with respect to source fields. Integrating out the electrons gives the effective action for the phonons
\begin{align}
	&e^{i S_{\rm eff}[X ]}
	=
	e^{i S_{x}[X] + i\Gamma[X]},\,\,\,
	\\
	&\Gamma[X]=-i\log\Big\langle
	e^{iS_{cx}[\bar c,c,X]}
	\Big\rangle_{cc},
\end{align}
where $\langle\cdots\rangle_{cc}=\frac{1}{Z_{cc}}\int\mathcal{D}[\bar c,c] e^{iS_{cc}[\bar c,c]} \cdots$ is the bare electronic expectation value. In order to proceed towards stochastic equations, we represent the fields $X_j$ in terms of their so-called quantum and classical components, which are functions of the physical time $t\in(-\infty,\infty)$. We first denote by $X^{\pm}$ the fields  on the upper/lower branch as a function of the physical time $t_j$ $(j=1,...,N)$,
\begin{align}
	\label{Xpm}
	&X^{+}_j \equiv X(t_j^+) = X_{j},  
	\\
	\label{Xpm2}
	&X^{-}_j \equiv X(t_j^-) = X_{2N+1-j}. 
\end{align}

Quantum and classical components are introduced via the Keldysh rotation, 
\begin{align}
\begin{pmatrix}
X^\text{cl}_j \\
X^\text{q}_j
\end{pmatrix} = \frac{1}{2} 
\begin{pmatrix}
X^+_j + X^-_j \\
X^+_j - X^-_j
\end{pmatrix} .
\end{align}
The classical component of the field $X^\text{cl}$ represents the average of the fields on the upper and lower branches of the contour, while the quantum component $X^\text{q}$ is the semi-difference of the two. In a purely classical configuration, we expect the fields on the two branches to be equal, i.e. $X^+_j = X^-_j$, leading to $X^\text{q}_j = 0$, which explains the notation~\cite{Sieberer2016}. 
With this, the free-phonon part of the action can be recast in the form
\begin{align} \label{action_x_clq1}
	S_x =
	& - 2 \delta_t \sum_{j=2}^{N-1} X_j^{\text{q}} \big( \ddot{X}_j^{\text{cl}} + \Omega^2 X_j^{\text{cl}} \big) 
	+
	b.t.,
\end{align}
with the discrete second derivative
\begin{align}
	\ddot{X}_j^{\text{cl}} = \frac{X_{j+1}^{\text{cl}} - 2 X_j^{\text{cl}} + X_{j-1}^{\text{cl}}}{\delta_t^2} \;.
\end{align}
The abbreviation $b.t.$ denotes terms involving only the fields $X^{\pm}_{1}$, $X^{\pm}_{2}$ $X^{\pm}_N$, $X_{N-1}^{\pm}$ on the boundary of the contour, which will not be important in the following. (They set the initial condition for the stochastic equation at $t\to-\infty$, which becomes irrelevant because the system evolves in the presence of a damping term, see below.) The electron-phonon action \eqref{action_cx1} becomes
\begin{align} \label{action_cx_clq}
	S_{cx} = 
	& - \delta_t \sqrt{2\Omega} g \sum_{j=2}^{N-1} 
	\big( O^+_{j+1} - O^-_{j-1} \big) X^{\text{cl}}_j 
	\nonumber \\
	& 
	- \delta_t \sqrt{2\Omega} g \sum_{j=2}^{N-1}  
	\big( O^+_{j+1} + O^-_{j-1} \big) X^{\text{q}}_j \;
	+b.t.,
\end{align}
where in analogy to Eqs.~\eqref{Xpm} and \eqref{Xpm2} we use the notation $O^+_{j}\equiv O(t_j^+)=O_j$, and $O^-_j\equiv O(t_j^-)=O_{2N+1-j}$.

The effective potential $\Gamma[X]$ is expanded in a Taylor series in the quantum variable
\begin{align}
	&\Gamma[X^{\text{cl}},X^{\text{q}}]=\sum_{n=0}^\infty \frac{1}{n!} \sum_{j_1,...,j_n} X^{\text{q}}_{j_1}\cdots X^{\text{q}}_{j_n} 
	\,\tilde \Pi_{j_1,...,j_n},
	\label{seriesgamma}
	\\
	&\tilde \Pi_{j_1,...,j_n}=\frac{-i\,\partial^n }{\partial X^{\text{q}}_{j_1}\cdots  \partial X^{\text{q}}_{j_n}}\log\Big\langle
	e^{iS_{cx}[\bar c,c,X]}
	\Big\rangle_{cc} \Big|_{X^{\text{q}}=0}.
\end{align}
To interpret the coefficients $\tilde \Pi$, we define the action
\begin{align}
	\label{glacte}
	S_{cl}
	=
	S_{cc}
	- \delta_t \sqrt{2\Omega}g \sum_{j=2}^{N-1} \big(O^+_{j+1} -O^-_{j-1} \big)X^{\text{cl}}_j +b.t. .
\end{align}
This describes a purely electronic model where electrons at the impurity are subject to a fluctuating field
\begin{align}
	\label{fluct}
	\hat H_X(t) = \sqrt{2\Omega}g X^{\text{cl}}(t)\, \hat O
\end{align}
in the Hamiltonian, with the time-dependent force $\sqrt{2\Omega}g X^{\text{cl}}(t)=\sqrt{2\Omega}g X^{\text{cl}}_j \text{~for~} t\in[t_j,t_{j+1}]$. With this,
\begin{align}
	\tilde \Pi_{j_1,...,j_n}&\equiv (-i)^{n+1}(2\sqrt{2\Omega}g\delta_t)^n   \langle  \bar O_{j_1} \cdots \bar O_{j_n}\rangle_{cl}^\text{con},
\end{align}
where $\langle\cdots\rangle_{cl}^\text{con}$ are the connected correlation functions for the electrons in  the presence of the fluctuating force \eqref{fluct}, and $\bar O_{j}= (O^+_{j+1}+O^-_{j-1})/2$. In particular
\begin{align}
	& \tilde \Pi_{j} = - 2\sqrt{2\Omega}g \delta_t \langle \bar O_{j}\rangle_{cl},
	\\
	\label{p2}
	& \tilde \Pi_{j,l} =  -\delta_t^24\Omega g^2
	(\chi^K_{cl})_{j,l},
\end{align}
where 
\begin{align}
	\label{po3}
	(\chi^K_{cl})_{j,l}=-2i \big(\langle \bar O_j \bar O_l\rangle_{cl}-\langle \bar O_j \rangle_{cl}\langle\bar O_l\rangle_{cl}\big)
\end{align}
is recognized as the discrete representation of the Keldysh component
\begin{align}
	\chi^K_{cl}(t,t')
	&=
	-i\Big( \big\langle \hat O(t) \hat O(t') \big\rangle_{cl}^\text{con}
	+ t\leftrightarrow t'\Big)
	\label{keldysh}
\end{align}
of the connected $OO$ autocorrelation function of the electrons in  the presence of the fluctuating field term \eqref{fluct}.

\subsection{Semiclassical approximation}

The semiclassical approximation corresponds to truncating the expansion \eqref{seriesgamma} at the second order. The first order in the expansion \eqref{seriesgamma} is combined with the free phonon action \eqref{action_x_clq1} into 
\begin{align} 
	\label{action11}
	S_x +\Gamma^{(1)}
	=
	& -2 \delta_t\sum_{j=2}^{N-1} X_j^{\text{q}} \Big( \ddot{X}_j^{\text{cl}} + \Omega^2 X_j^{\text{cl}} +\sqrt{2\Omega}g \langle \bar O_{j}\rangle_{cl}\Big).
\end{align}
The quadratic term in Eq.~\eqref{seriesgamma}, $\Gamma^{(2)}[X]=-2\Omega \delta_t^2g^2 \sum_{j,l} (\chi^K_{cl})_{j,l}X^{\text{q}}_jX^{\text{q}}_l$, is decoupled using a Hubbard Stratonovich transformation with a real field $\xi$,
\begin{align}
	\label{hubbard_stratonovich01}
	e^{i\Gamma^{(2)}}&=
	e^{ -  \frac{1}{2} \delta_t^2 4\Omega \sum_{j,l} X^{\text{q}}_j \  (i g^2(\chi^K_{cl})_{ j,l})\ X^{\text{q}}_{l} }
	\nonumber\\
	& = \frac{1}{Z_\xi} \int\mathcal{D}[\xi]
	e^{ - \frac{1}{2} \sum_{j ,l} \xi_j A_{j,l} \xi_{l} + i 2\delta_t\sqrt{\Omega}  \sum_{j} \xi_j X^{\text{q}}_j },
\end{align}
with $(A^{-1})_{j,l}=i g^2  (\chi^K_{cl})_{j,l}$. Here $\int\mathcal{D}[\xi] = \int \prod_{j}d \xi_j$, and 
\begin{align}
	Z_{\xi}=\int\mathcal{D}[\xi] e^{-\frac{1}{2} \sum_{j , l} \xi_j A_{j,l} \xi_{l}}
\end{align}
is the normalization factor. The Gaussian integral is convergent, because $(i\chi^K_{cl})_{j,l} $ is a positive definite matrix [c.f., Eq.~\eqref{po3}].
Combining all terms which are linear in $X^{\text{q}}$ in Eqs.~\eqref{action11} and \eqref{hubbard_stratonovich01} gives a factor $e^{-i 2\delta_tX_{j}^{\text{q}} F_j}$, with 
\begin{align}
	F_j=\ddot{X}_j^{\text{cl}} + \Omega^2 X_j^{\text{cl}} +\sqrt{2 \Omega}g \langle \bar O_{j}\rangle_{cl} -\sqrt{\Omega}\xi_j.
\end{align}
The integral over $X^{\text{q}}$ is then  performed analytically,
\begin{align}
	\int dX^{\text{q}}_j e^{-i 2\delta_tX_{j}^{\text{q}} F_j} = \frac{ \pi}{\delta_t} \delta(F_j).
\end{align}
With this, the path integral has been reduced to (up to  constants)
\begin{align}
	Z
	=
	\int\mathcal{D}[\xi] 
	\int\mathcal{D}[X^{\text{cl}}]
	\frac{1}{Z_{\xi}} 
	e^{-\frac{1}{2}\sum_{j,l}\xi_j A_{j,l} \xi_{jl}} \prod_{j}\delta(F_j).
	\label{5637hhh}
\end{align}
The  term $\prod_{j}\delta(F_j)$ constrains the values $X^\text{cl}_{j}$ to a trajectory which satisfies at each time step $j\ge 2$ the (discretized) equation of motion
\begin{align}
	\label{eom}
	\ddot{X}_j^{\text{cl}}=-\Omega^2 X_j^{\text{cl}} -\sqrt{2 \Omega}g \langle \bar O_{j}\rangle_{cl} +\sqrt{\Omega}\xi_j
\end{align}
which can be rewritten in the form of two first order differential equations
\begin{align}
\label{seom1} 
&V_{j} = V_{j-1}+\delta_t \big(-\Omega^2 X^{\text{cl}}_j -\sqrt{2\Omega }g\langle \bar O_{j}\rangle_{cl} +\sqrt{\Omega}\xi_j\big),
\\
\label{seom2} 
&X^{\text{cl}}_{j+1}=X^{\text{cl}}_{j}+\delta_t V_j.
\end{align}
We point out that if the expansion in the quantum variable in Eq.~\eqref{seriesgamma} is truncated at the leading order, the $\xi$ integral will be absent, and Eq.~\eqref{eom} becomes the equation of motion obtained from a mean-field decoupling of the electron-phonon interaction, which is reminiscent of the result from Ehrenfest dynamics~\cite{tenBrink2022}. On the other hand, in the derivation of Eq.~\eqref{eom}, we did not make any assumption on the separation of the time-scales between the electronic and the lattice subsystems, so this equation cannot be regarded as a Born-Oppenheimer approximation. The term $\xi$ can be viewed as a stochastic force, whose statistics is determined by the matrix $A$, which itself depends on the trajectory $X^\text{cl}_j$. With this, the path integral \eqref{5637hhh} defines a nonlinear non-Markovian stochastic differential equation. A scheme for the solution of this problem is provided in App.~\ref{nonmark}.

\subsection{White noise limit}

The stochastic equations can be further simplified if the electronic timescale $\tau_e$, which is determined by the bandwidth of the bath, is much shorter than the period $1/\Omega$ of the phonon. In this case, we can try to numerically solve for the phonon dynamics on a time grid $\Delta t$ which is sufficiently short compared to $1/\Omega$, but still long compared to $\tau_e$, $\tau_e\ll \Delta t \ll 1/\Omega$. Electronic correlation functions, in particular the autocorrelation function \eqref{keldysh}, vanish for  time differences $t-t'\gg\tau_e$, which implies that also the noise becomes uncorrelated on these times,
\begin{align}
	\label{llwkfwf}
	\langle
	\xi(t_1) \xi(t_2)
	\rangle \to 0 \text{~~~for~~} |t_1-t_2|\gg \tau_e.
\end{align}
As a consequence, the phonon dynamics on the coarser time grid $\Delta t$ should be reduced to a stochastic equation with a force that is uncorrelated between different time steps (white noise). To derive this equation, let us formally integrate Eqs.~\eqref{seom1}  and \eqref{seom2} over a time interval of duration $\Delta t$ (to simplify the notation, we set the initial time of the interval to $0$), 
\begin{align}
	&X^{\text{cl}}(\Delta t) =X^{\text{cl}}_{\rm 0}(\Delta t)+\frac{1}{\Omega} \int_0^{\Delta t} \!\!\!\!d\bar t  \sin[\Omega (\Delta t- \bar t )]f(\bar t),
	\\
	&V(\Delta t) = V_{\rm 0}(\Delta t) +\int_0^{\Delta t} \!\!\!d\bar t \cos[\Omega(\Delta t-\bar t )] f(\bar t).
\end{align}
Here $f(t)=-g\sqrt{2\Omega}\langle \bar O(t)\rangle_{cl} +  \sqrt{\Omega}\xi(t)$ is the force, and $X_{\rm 0}(t)= X(0)\cos(\Omega t) + \Omega^{-1}V(0)\sin(\Omega t)$ and $V_{\rm 0}(t)= V(0)\cos(\Omega t) -\Omega X(0)\sin(\Omega t)$ correspond to the solution of the free oscillator. For sufficiently short times $ \Delta t\ll 1/\Omega$, the increment of the position is therefore 
\begin{align}
	\label{xeheje}
	&X^{\text{cl}}(t) \approx  X^{\text{cl}}(0)+ V(0)t + \cdots,
\end{align}
where the omitted terms are of order $\mathcal{O}(t^{3/2})$. To see the latter, consider the correction to Eq.~\eqref{xeheje}, $\Delta X_f = \frac{1}{\Omega} \int_0^t dt_1 \sin[\Omega(t-t_1)] f(t_1) =   \int_0^t d t_1 \, t_1 f(t_1) +\mathcal{O}(t^2)$, which is a random variable with variance  $\sigma_{\Delta X_f}^2=\int_0^t d t_1 dt_2 \, t_1 t_2 \langle f(t_1) f(t_2)\rangle $. Even for a delta-correlated force term $\kappa(t_1,t_2)=\langle f(t_1) f(t_2) \rangle\sim \delta(t_1-t_2)$  (white noise) this gives just $\sigma_{\Delta X_f} \sim t^{3/2}$; for a smooth function $\kappa(t_1,t_2)=\langle f(t_1) f(t_2) \rangle$ the square root of the variance is $\sigma_{\Delta X_f} \sim t^2$. Similarly, the increment of $V$ is 
\begin{align}
	&V(\Delta t)-V(0) =  -\Omega^2 X(0)\Delta t + \int_0^{\Delta t} dt_1 f(t_1),
	\label{llelelele}
\end{align}
up to terms of $\mathcal{O}(\Delta t^2)$. We first analyze the contribution of $\langle\bar O(t) \rangle_{cl}$ to the force integral,
\begin{align}
	\label{gegheje}
	\Delta V_O = - \sqrt{2\Omega}\,g \int_0^{\Delta t} d t_1 \langle\bar O(t_1) \rangle_{cl}.
\end{align}
The variable $ \langle\bar O(t_1) \rangle_{cl}$ depends on the trajectory $X^{\text{cl}}(t)$ during the interval $[0,\Delta t]$. However, since  $X^{\text{cl}}(t)-X^{\text{cl}}(0)$ remains small [c.f.~Eq.~\eqref{xeheje}], we can obtain $\langle\bar O(t_1) \rangle_{cl}$ by the linear response in the difference $\Delta X(t)=X^{\text{cl}}(t)-X^{\text{cl}}(0)$,
\begin{align}
	\label{aeffe}
	\langle\bar O(t_1) \rangle_{cl} = \langle\bar O(t_1) \rangle_{cl,0} + \sqrt{2\Omega}g\int_0^{t_1} \!\!\!dt_2 \, \chi^R_{cl,0}(t_1,t_2) \Delta X(t_2).
\end{align}
Here $\langle\bar O(t_1) \rangle_{cl,0}$ is the expectation value calculated for the electronic model \eqref{glacte} in which the field  $X^{\text{cl}}(t)$  is frozen to  $X^{\text{cl}}(0)$ for $t>0$. Accordingly, $\chi^R_{cl,0}(t_1,t_2)$ is the response function 
\begin{align}
	\label{retarder}
	\chi^R_{cl}(t_1,t_2)=-i\theta(t_1-t_2) \langle [\hat O(t_1),\hat O(t_2)] \rangle_{cl}
\end{align}
in that model. When Eq.~\eqref{aeffe} is inserted into \eqref{gegheje}, we can to leading order replace $\langle\bar O(t_1) \rangle_{cl,0}=\langle\bar O(0) \rangle_{cl}$, and $\Delta X(t_2)$ by \eqref{xeheje},
\begin{align}
	\Delta V_O &= - \sqrt{2\Omega}g \Delta t \langle\bar O(0) \rangle_{cl}
	\nonumber
	\\
	&-
	V(0) 2 g^2\Omega  \int_0^{\Delta t} d t_1 
	\int_0^{t_1} d t_2 
	\,\chi^R_{cl,0}(t_1,t_2)t_2.
	\label{ddqw;w;}
\end{align}
The function $\chi^R_{cl,0}(t_1,t_2)$ decays to zero when the time difference $t_1-t_2$ is much larger than the electronic timescale $\tau_e$. With the separation of timescales  $\tau_e \ll \Delta t \ll 1/\Omega$, the  double integral in Eq.~\eqref{ddqw;w;} then becomes (substituting $t_2=t_1-s$)
\begin{align}
	&\int_0^{\Delta t} d t_1 
	\int_0^{t_1} ds\,
	\chi^R_{cl,0}(t_1,t_1-s) (t_1-s)
	\nonumber
	\\
	&\approx
	\int_0^{\Delta t} d t_1 
	\int_0^{\infty} ds\,
	\chi^R_{cl,0}(t_1,t_1-s) (t_1-s)
	\\
	&=
	\int_0^{\Delta t} d t_1 
	(t_1+i\partial_\omega) \chi^R_{cl,0}(t_1,\omega) |_{\omega=0},
	\label{almostmadeit}
\end{align} 
with the backward Fourier transform
\begin{align}
	\chi^R_{cl,0}(t,\omega)=
	\int_0^{\infty } ds \,e^{i\omega s} \chi^R_{cl,0}(t,t-s).
\end{align}  
When replacing Eq.~\eqref{almostmadeit} with its expression to leading order in $\Delta t$, Eq.~\eqref{ddqw;w;}  becomes
\begin{align}
	\label{forceetrem}
	\Delta V_O &= \Delta t\,\big(- \sqrt{2\Omega}g \langle\bar O(0) \rangle_{cl} -\Omega\Gamma(0)V(0)\big),
\end{align}
with
\begin{align}
	\Gamma(t) &=-2g^2 \partial_\omega\text{Im}\chi^R_{cl}(t,\omega) |_{\omega=0}.
\end{align}
Here, we also used the anti-hermitian symmetry of the correlation functions in time.

Next, we analyze the contribution from the stochastic term to the force integral in Eq.~\eqref{llelelele},
\begin{align}
	\label{gegheje01}
	\Delta V_\xi =\sqrt{\Omega} \int_0^{\Delta t} d t_1 \xi (t_1).
\end{align}
The noise $ \xi (t_1)$ is determined by the Keldsyh matrix $\chi^K_{cl}(t_1,t_2)$ for $t_1,t_2 \le t$. To leading order in $\Delta t$,  we can replace the latter by the correlation function $\chi^K_{cl,0}$, obtained again by the electronic model \eqref{glacte} in which the field  $X^{\text{cl}}(t)$  is frozen to  $X^{\text{cl}}(0)$ for $t>0$. In the limit of $\tau_e\to 0$ [with Eq.~\eqref{llwkfwf}],  the  noise therefore depends only on the Keldysh matrix in the interval $[0,\Delta t]$, and becomes a statistically independent variable with variance
\begin{align}
	\sigma_{\Delta V_\xi}^2
	&= \Omega \int_0^{\Delta t}  d t_1 \int_0^{\Delta t} d t_2  \langle  \xi (t_1) \xi (t_2)\rangle
	\\
	&\approx \Omega  \int_0^{\Delta t}  d t_1 \int_0^{\Delta t} d t_2  ig^2\chi^K_{cl,0}(t_1,t_2)
	\\
	\label{line59}
	&\approx \Omega\int_0^{\Delta t}  d t_1 \int_{-\infty}^{+\infty} \!d t_2\, ig^2\chi^K_{cl,0}(t_1,t_1-s)
	\\
	&= \Omega\int_0^{\Delta t}  d t_1  ig^2\chi^K_{cl,0}(t_1,\omega)|_{\omega=0} 
	\\
	&\approx -\Delta t \,\Omega\,g^2\text{Im}\chi_{cl}^K(0,\omega)|_{\omega=0} .
\end{align}
In the third step \eqref{line59} we have made use of the decay of $\chi^K_{cl,0}(t_1,t_2)$ on times shorter than $\Delta t$, and the last step is again correct to leading order in $\Delta t$.

In summary,  Eqs.~\eqref{seom1} and \eqref{seom2} in the limit $\tau_e\to0$ reduce to the stochastic equations
$V_{j} = V_{j-1}+\Delta t F_{j-1}$ and $X^{\text{cl}}_{j+1}=X^{\text{cl}}_{j}+\Delta t V_{j}$, with the force
\begin{align} \label{force}
	F_j=-\Omega^2 X^{\text{cl}}_j -g\sqrt{2\Omega} \langle \bar O_{j}\rangle_{cl} - \Omega \Gamma_j V_{j}+\sqrt{\Omega}\xi_j
\end{align}
that contains a damping term 
\begin{align}
	\Gamma_j =
	-2 g^2 \partial_\omega \text{Im}\chi^R_{cl}(t_j,\omega) |_{\omega=0},
\end{align}
and a white noise $\langle \xi_j\rangle=0$, $\langle \xi_j\xi_{j'}\rangle = K_j \delta_{j,j'} \Delta t^{-1} $ with 
\begin{align}
	K_j=-g^2\text{Im}\chi_{cl}^K(t_j,\omega)|_{\omega=0}.
\end{align}
The stochastic evolution of the phonons must be solved together with a time propagation of the electron dynamics in the presence of the time-dependent field \eqref{fluct}, which determines the force $\sqrt{\Omega} \langle \bar O_{j}\rangle_{cl}$, the damping $\Gamma_j$, and the noise power $K_j$. In equilibrium, i.e., for time-independent $X^{\text{cl}}$, the correlation function $\chi_{cl}^K(t,\omega)$ [Eq.~\eqref{keldysh}] and  $\chi_{cl}^R(t,\omega)$ [Eq.~\eqref{retarder}] do not depend on time and satisfy the fluctuation dissipation relation 
\begin{align}
	\chi_{cl}^K(\omega)=2i\coth\Big(\frac{\omega\beta}{2}\Big) \text{Im}\chi_{cl}^R (\omega).
\end{align}
Hence we have $K=2\Gamma T$, i.e, the standard Einstein relation between noise and damping.

In the end, we point out that, despite the apparent similarity of our treatment with the better known Born-Oppenheimer approximation, the stochastic semiclassical theory does not rely on an adiabatic assumption for the electronic degree of freedom. Thus it does not have the same limitations as the Born-Oppenheimer approximation, 
e.g. in the presence of level crossings in the electronic band structure. Nevertheless, we emphasize that the separation of time scales between the electronic and the phononic subsystems is a relevant assumption leading to Eq.~\eqref{force}.

\subsection{ \label{sec:QBE} Quantum Boltzmann equation for the electronic problem}
After the quantum phonon has been replaced with the semiclassical stochastic one, $X^\text{cl}$, one still needs to solve the electron impurity model with action \eqref{glacte}, with a time-dependent force  [c.f.~Eq.~\eqref{fluct}]. In the limit of well-separated timescales for the electrons and the lattice, the  time evolution of the local electronic system on the impurity can be obtained from a non-perturbative  quantum Boltzmann equation (QBE) for the distribution function $\mathcal{F} ( \omega, t ) = G^< (\omega, t)/(2 \pi i A (\omega,t))$, which is given by the ratio between the lesser component of the electronic Green's function  $G^<$ (occupied density of states), and the spectral function $A(\omega,t)=-\frac{1}{\pi} \Im G^R(\omega,t)$ \cite{Picano2021}. Dynamical quantities depending on time and frequency are understood in terms of the Wigner transform of two time functions
\begin{align}
Y(\omega,t) = \int ds \,e^{i\omega s} Y(t+s/2,t-s/2)
\end{align}
with respect to relative time $s$ at given average time $t$. The QBE  provides an equation of motion for both the distribution and the spectral function. The basic assumption is that the evolution of electronic spectra and distribution functions with average time $t$ is much slower than $1/ \delta \omega$, where $\delta \omega$ is given by the bandwidth of the relevant spectral features \cite{Picano2021}.  The QBE then gives an equation for the evolution of the distribution function, $\partial_t \mathcal{F} ( \omega, t ) = I_{\omega} [\mathcal{F}]$, with scattering integral
\begin{align} \label{scatt_int}
I_{\omega} [\mathcal{F}] = & -i \Gamma^<( \omega, t )  + 2i \mathcal{F} ( \omega, t ) \,\text{Im}\, \Gamma^R (\omega, t),
\end{align}
with the self-energy $\Gamma(\omega, t)$. For the Anderson Holstein model (see below), $\Gamma(\omega, t)=\Delta(\omega, t)$ with $\Delta$ being the hybridization of the impurity with the noninteracting bath. More generally, additional self-energy contributions $\Sigma_\text{int}$ have to be included due to interactions, which however can also be assumed to be a functional of the full Green's function $G$, $\Sigma_\text{int}=\Sigma_\text{int}[G]$. In this case, $\Gamma$ reads $\Gamma(\omega, t)=\Delta(\omega, t) + \Sigma_\text{int}(\omega, t)$. The self-energy and the spectrum that appear in Eq.~\eqref{scatt_int} are evaluated non-perturbatively, by solving at each timestep an auxiliary steady-state problem which is designed to have the same interaction, and a non-equilibrium steady state distribution function $\mathcal{F}_{NESS}(\omega)=\mathcal{F}(\omega,t)$, which, for a lattice problem, can be computed with DMFT (for details, see Ref.~\onlinecite{Picano2021}). 
The assumption here is that the electronic structure (spectral function) is determined by the electronic distribution function at the same time (the so-called \textit{instantaneous response approximation} in Ref.~\onlinecite{Picano2021}). We emphasize that the computational effort, in the presence of a nonzero local interaction (such as a Hubbard $U$, which we do not consider in the following), would depend on the specific impurity solver used, but would not dramatically change as long as the white noise approximation remains valid.

\section{\label{sec:holstein_model}Holstein impurity model}

\subsection{\label{sec:model} Model}
The quantum impurity model considered here is represented as a Hamiltonian, $H_{QI}$, with three basic terms: $H_\text{imp}$, which describes the impurity, $H_\text{res}$, the infinite noninteracting system with continuous spectrum to which the impurity is coupled (so-called bath or reservoir), and $H_\text{coupl}$, the coupling term between the impurity and the bath: 
\begin{align} 
	\label{hamilt}
	H_{QI}=H_\text{imp}+H_\text{res}+H_\text{coupl}.
\end{align}
The Hamiltonian of the impurity is
\begin{align}
& H_\text{imp}=H_\text{x}+H_\text{cx}+H_\text{cc} \nonumber \\
			  & = \frac{1}{2} \Big( \Omega^2 \hat X^2 + \hat P^2\Big) + \sqrt{2 \Omega}g \hat{X} (\sum_{\sigma}n_\sigma -1) +H_\text{cc} . 
\label{hamilt_imp}
\end{align}
The first term corresponds to Eq.~\eqref{Hx}, with quadratures $\hat X = (\hat d + \hat d^\dagger)/ \sqrt{2 \Omega}$ and $\hat P = i \sqrt{\Omega / 2} (\hat d^\dagger - \hat d)$ expressed in terms of phononic creation and annihilation operators $\hat d^\dagger$ and $\hat d$; the second one is given in Eq.~\eqref{Hcx}, where now we set the operator $\hat{O}=(\sum_{\sigma}n_\sigma -1)$. $H_\text{cc}$ is the purely electronic part of the Hamiltonian, that can include, e.g., the chemical potential or the on-site Hubbard interaction, $H_\text{cc}=-\mu \sum_{\sigma}n_\sigma + Un_\uparrow n_\downarrow$. However, in the simulations we set both $\mu$ and $U$ to zero, 
meaning that the impurity is at half-filling.

The terms $H_{\text{res}}+H_{\text{coupl}}$ in Eq.~\eqref{hamilt} define a quadratic Hamiltonian that describes the hybridization of the impurity with a reservoir (bath) characterized by the Green's function $g_{p,p'}(t , t')$, with $p$ and $p'$ labels for the energy levels of the reservoir. The effect of the bath is included through a self-energy correction to the impurity Green's function. The hybridization function reads 
\begin{align}
\label{Delta}
\Delta(t,t')=\sum_{p,p'} V_{0,p} g_{p,p'}(t,t') V_{p',0},
\end{align}
where $V_{0,p}$ is the hopping integral from the impurity to the bath level $p$.
The bath is assumed to be always in equilibrium at temperature $T=\beta^{-1}$, 
and it is therefore convenient to represent the self-energy in real frequency space $\Delta (\omega)$. In this case, one  
needs to specify the
bath
spectrum
only, 
$\mathcal{A}_{\Delta}(\omega)=-\frac{1}{\pi}\text{Im}\Delta^R(\omega+i0)$, 
and 
the lesser component is given by the fluctuation-dissipation relation $\Delta^<(\omega)=-2if_\beta(\omega) \Im \{\Delta^R(\omega)\}$, where $f_\beta$ is the Fermi distribution at inverse temperature $\beta$.
In our implementation, the density of states of the bath is assumed to be semicircular leading to a bath spectral function
\begin{align}
\label{spectrum_Delta}
\mathcal{A}_{\Delta}(\omega)=  V^2\frac{2}{\pi D^2}  \theta (D^2-\omega^2) \sqrt{D^2-\omega^2},
\end{align}
with $\theta (\omega)$ the Heaviside step function and $D$ the half bandwidth. For all simulations below, we set $D=2$ to define the unit of energy, while the coupling $V$ between the impurity and the bath is chosen to be $V=0.1581 D$. We stress that the bath plays a crucial role in transmitting the information on the temperature to the semiclassical phonons.

\subsection{\label{sec:MonteCarlo} Monte Carlo method}

To test the semi-classical theory we will provide benchmarks against established methods for the Holstein impurity problem. 
The Holstein model at temperature $T>0$ can be solved numerically exactly (within error bars) using QMC methods. Here we consider the hybridization-expansion CTQMC approach developed in Ref.~\onlinecite{Werner2007}. This method performs a stochastic expansion of the partition function in powers of the hybridization function $\Delta$. At expansion order $n_\sigma$ for spin $\sigma$, the $n_\sigma!$ diagrams corresponding to a given imaginary-time sequence of fermionic creation and annihilation operators can be summed up into a determinant of a matrix $M_\sigma^{-1}$ \cite{Werner2006}, so that the weight of the Monte Carlo configuration can be expressed as
\begin{align}
	w(\{\psi_i(\tau_i)\}) =& \,\text{Tr}_c\Big\langle T_\tau e^{-\int_0^\beta H_\text{imp}(\tau)} \psi_{2n}(\tau_{2n})\ldots \psi_1(\tau_1)\Big\rangle_b \nonumber\\
	& \times d\tau_1\ldots d\tau_{2n}\prod_\sigma (\text{det} M_\sigma^{-1})s_\sigma,
	\label{eq_w}
\end{align}
where the $\psi_i(\tau_i)$ denote the (time-ordered) electron creation or annihilation operators and $s_\sigma$ is 1 ($-1$) if the spin-$\sigma$ operator with the lowest time argument is a creation (annihilation) operator. To evaluate the phonon expectation value, a Lang-Firsov transformation \cite{Lang1962} is introduced to shift $X\rightarrow X-X_0$, with $X_0=(\sqrt{2}g/\Omega^{\frac{3}{2}})(\sum_\sigma n_\sigma-1)$. This transformation introduces the polaron operators $\tilde c_\sigma^\dagger=e^{\frac{g}{\Omega}(d^\dagger-d)}c^\dagger_\sigma$, $\tilde c_\sigma=e^{-\frac{g}{\Omega}(d^\dagger-d)}c_\sigma$ and decouples the electrons from the phonons in the transformed impurity Hamiltonian. The transformation also shifts the local interaction and chemical potential as $U\rightarrow \tilde U=U-2\frac{g^2}{\Omega}$ and $\mu \rightarrow \tilde \mu=\mu-\frac{g^2}{\Omega}$. One may then integrate out the phonons, to obtain a phonon-related weight factor 
\begin{align}
	&w_b(\{\psi_i(\tau_i)\})=\text{exp}\Bigg[ -\frac{g^2/\Omega^2}{e^{\beta\Omega}-1}\Big(n(e^{\beta\Omega}+1)\Big)\nonumber\\
	& \quad +\sum_{2n\ge i>j\ge 1} s_i s_j\big\{ e^{\Omega(\beta-(\tau_i-\tau_j))}+e^{\Omega(\tau_i-\tau_j)} \big\}\Big)\Bigg],
\end{align}
where $s_i=1$ ($-1$) if the $i$th operator is a creation (annihilation) operator. The weight (\ref{eq_w}) becomes the product of this bosonic weight factor and the weight corresponding to a usual Anderson impurity model (without phonons, but with the modified $\tilde U$, $\tilde \mu$):
\begin{align}
	& w(\{\psi_i(\tau_i)\}) = w_b(\{\psi_i(\tau_i)\}) \tilde w_\text{AIM}(\{\psi_i(\tau_i)\}).
\end{align}
Based on these weights, one then samples all possible diagrams using Monte Carlo updates which insert/remove pairs of creation and annihilation operators, or which shift the positions of the operators \cite{Werner2006}. 

To measure the phonon distribution function $P(X)$, we calculate the expectation values 
\begin{align}
\label{paa}
p (a) &= \Big\langle \cos \Big( a \sqrt{\Omega} X \Big) \Big\rangle_\text{MC}   \\
&= \int dX \ P (X) \cos \Big( a \sqrt{\Omega} X \Big) \nonumber  
\end{align}
for different $a$. To derive the measurement formula, we first discuss the measurement of $\langle e^{ia \sqrt{\Omega} X}\rangle_\text{MC}$. This measurement formula is obtained by inserting the operator $e^{ia \sqrt{\Omega} X}$ at $\tau=0$ into the expression (\ref{eq_w}), which defines $w^X(\{\psi_i(\tau_i)\})=\text{Tr}_c\Big\langle T_\tau e^{-\int_0^\beta H_\text{imp}(\tau)} \psi_{2n}(\tau_{2n})\cdots \psi_1(\tau_1) e^{ia \sqrt{\Omega} X} \Big\rangle_b d\tau_1\cdots d\tau_{2n}$ $\times\prod_\sigma (\text{det} M_\sigma^{-1})s_\sigma$. During the Monte Carlo sampling, we then measure the ratio $w^X(\{\psi_i(\tau_i)\})/w(\{\psi_i(\tau_i)\})$. Since the additional $e^{ia \sqrt{\Omega} X}$ factor only modifies the bosonic factor, this amounts to measuring the ratio $w_b^X(\{\psi_i(\tau_i)\})/w_b(\{\psi_i(\tau_i)\})$, where $w_b^X(\{\psi_i(\tau_i)\})$ is the bosonic weight factor obtained with the additional operator $e^{ia \sqrt{\Omega} X}$ at $\tau=0$. This ratio can be expressed as
\begin{align}
	&\frac{w_b^X(\{\psi_i(\tau_i)\})}{w_b(\{\psi_i(\tau_i)\})} = \exp\Bigg[-\frac{a^2}{4}\frac{e^{\beta\Omega}+1}{e^{\beta \Omega}-1} -ia \sqrt{\Omega}X_0(\tau=0) \Bigg] \nonumber \\
	& \!\times\exp\Bigg[ -\frac{i}{e^{\beta\Omega}-1}\sum_j s_j\frac{g}{\Omega}\frac{a}{\sqrt{2}}\big( e^{\Omega(\beta-\tau_j)}-e^{\Omega\tau_j}\big) \Bigg].
\end{align}
Note that because of the Lang-Firsov shift, this expression depends on $X_0(\tau=0)$ and hence on the occupation of the impurity at $\tau=0$ in the measured configuration. Since the first factor is independent of the Monte Carlo configuration, we can express the measurement formula for $p(a)$ [Eq.~\eqref{paa}]
as
\begin{align}
	&\Big\langle \cos(a \sqrt{\Omega} X) \Big\rangle_\text{MC} = \exp\Bigg[-\frac{a^2}{4}\frac{e^{\beta\Omega}+1}{e^{\beta\Omega}-1}\Bigg] \nonumber \\
	& \times \Bigg\langle \cos\Bigg[ a \sqrt{\Omega} X_0(\tau=0)\nonumber\\
	&\hspace{4mm}+\frac{1}{e^{\beta\Omega}-1}\sum_j s_j\frac{g}{\Omega}\frac{a}{\sqrt{2}}\big( e^{\Omega(\beta-\tau_j)}-e^{\Omega\tau_j}\big) \Bigg]\Bigg\rangle_\text{MC}. 
\end{align}
We measure $p(a)$ on a sufficiently fine $a$-grid ($\Delta a=0.02$ in the calculations below) and then compute the phonon distribution function by inverting Eq.~\eqref{paa} as
\begin{align}
	\label{postprocess}
	P(X)=\frac{1}{2\pi}\int da \, p(a) \cos(a \sqrt{\Omega} X). 
\end{align} 
Some tests of this measurement scheme against ED for a model with a single bath site are shown in Appendix~\ref{sec:QMC_ED}.

\subsection{\label{sec:MigdalApprox} Second-order Migdal approximation} 
This section discusses the solution of the Anderson-Holstein impurity model with the self-consistent Migdal approximation \cite{Randi2017}. In the Migdal approximation, one treats within a Gaussian approximation the quantum fluctuations of the phonons around the average order parameter, represented by the displacement $\langle X \rangle$ of the atoms. 
The self-consistent Migdal approximation is the lowest order approximation for the electronic self-energy that allows to treat the renormalization of the phonons induced by the electron-phonon coupling: The vibrational mode evolves as a consequence of the interaction with the electrons and, in turn, influences the electronic dynamics; the electrons couple back to the phonons in the form of a phonon self-energy (polarization bubble). The expectation values that will be shown in the following denote expectation values of quantum operators. The Dyson equation for the impurity Green's function reads
\begin{align}
	\label{Dyson_G}
	&[i \partial_t + \mu -h_\text{loc}(t)]G(t,t') \nonumber \\
	&- [\Delta(t,t')+\Sigma_\text{el-ph}^{\text{II}}(t,t')] * G(t,t') =\mathcal{\delta}_{\mathcal{C}}(t,t')
\end{align} 
in the Keldysh formulation  (following the notation for two-time Green's functions in Ref.~\cite{Aoki2014}). We are considering here the spin-symmetric phase and omitting spin indices in  $\Sigma$, $G$ and $\Delta$.   The term $h_\text{loc}$ defines a  time-local (Hartree) contribution to the electronic self-energy, i.e., a self-consistent on-site potential,
\begin{align} 
	\label{h_loc}
	h_\text{loc}(t)=\sqrt{2 \Omega}g \langle \hat X(t) \rangle.
\end{align}
Furthermore, we include the leading order self-consistent diagrammatic  corrections to the first order electronic self-energy $h_\text{loc}$  in  the  expansion  in  terms  of the fluctuations $\Delta \hat X(t) \equiv \hat X(t)- \langle \hat X(t) \rangle $.  
The second-order electronic self-energy, in the self-consistent Migdal approximation, is
\begin{align}
	\label{Sigma_Migdal}
	\Sigma_\text{el-ph}^{II}(t,t')= i g^2 G(t,t') D(t,t'),
\end{align}
where
\begin{align}
	\label{D_phonon}
	D(t,t')= -2i \langle T_{\mathcal{C}} \Delta \hat X(t) \Delta \hat X(t') \rangle.
\end{align}
To include the back-action of the electrons on the phonons on the same diagrammatic level, we include the phonon self-energy (polarization operator)
\begin{align}
	\label{Polariz}
	\Pi(t,t')=-2i g^2 \Omega G(t,t') G(t',t),
\end{align}
and solve the Dyson equation for the phonon propagator $D(t,t')$ in the form
\begin{align}
	\label{Dyson_D}
	[1- D_0(t,t')* \Pi(t,t')]* D(t,t') = D_0(t,t').
\end{align}
Here $D_0(t,t')$ is the non-interacting ($g=0$) phonon propagator,
\begin{align}
	\label{D0}
	D_0(t,t')
	=& -\frac{2 i}{\Omega} \cos [\Omega (t-t')]b_\beta (\Omega) \nonumber \\ 
	&- \frac{i}{\Omega} [ \theta_{\mathcal{C}}(t',t)e^{i \Omega (t-t')}+ \theta_{\mathcal{C}}(t,t')e^{-i \Omega (t-t')} ] ,
\end{align} 
where $b_\beta(\Omega)=1/(e^{\beta \Omega}-1)$ is the Bose 
function and $\theta_{\mathcal{C}}(t',t)$ is the Heaviside step function on the Keldysh contour $\mathcal{C}$. 

In the Migdal approximation,  the phononic distribution function $P_\text{start}(X)$ is Gaussian, thus it can be computed from the mean $\langle \hat X \rangle$ and the variance extracted from Eq.~\eqref{D_phonon}: $\text{Var}[X] \equiv \langle \Delta X^2 \rangle = \frac{i}{2} D^<(t,t)$. The equations can be solved numerically using the NESSi simulation package \cite{Schuler2020}.

\subsection{\label{sec:NCA} Non-crossing approximation}
In this section, we will present the solution of the single impurity Holstein model obtained within the non-crossing approximation (NCA)~\cite{Eckstein2010}. 
This scheme is based on a strong coupling expansion formulated in terms of pseudo-particles, which, in this case, have to take into account the mixed fermionic and bosonic nature of the local Fock space due to the presence of both bosons and fermions in the Hamiltonian \cite{Grandi2021b}. Within this approach, the local Fock space is composed of an electronic sector 
of dimension 
$4$ for a single band problem and a bosonic one of size $N_{\text{ph}}$ which defines the cutoff of the phononic subspace, so that $N_{\text{tot}} = 4 \times N_{\text{ph}}$. We set the highest energy bosonic state reachable in our simulation to $N_{\text{ph}} = 18$, similarly to what has been done is other contexts \cite{Grandi2021b,Sous2021}.

Given the perturbative nature of this method, we do not expect it to accurately describe the electronic properties of the system. However, it is interesting to compare to what extent it can provide a reasonable and consistent description of the phonon properties. Within NCA, we get access to the local reduced phonon density matrix $\rho^{\text{ph}} = \Tr_{\text{el}} \big[ \rho \big]$ from which we can build the phonon distribution function as 
\begin{align} \label{ph_distrib_NCA}
P (X) = \sum_{n,m = 0}^{N_{\text{ph}} - 1} \rho^{\text{ph}}_{n m} \varphi_n (X) \varphi_m (X) \;,
\end{align}
where
\begin{align}
\varphi_n (X) = N_n H_n (\sqrt{\Omega} X) e^{- \frac{\Omega X^2}{2}} 
\end{align}
is the $n$-th eigenfunction of the one-dimensional harmonic oscillator, $H_n (x)$ the $n$-th Hermite polynomial, and $N_n = \frac{1}{\sqrt{2^n n ! \sqrt{\pi}}}$ the corresponding normalization factor.

\subsection{\label{sec:semicl_eq_holstein} Semiclassical stochastic equations}

For completeness, we summarize in this section the semiclassical stochastic equations of Sec.~\eqref{sec:Semiclassical} for the particular case of the Anderson Holstein model. The equations of motion for the phonon read, in continuous time, $\dot{V} (t) = F (t)$, and 	$\dot{X}^{\text{cl}} (t) = V (t)$,  with the force field
\begin{align} \label{force_holstein}
	F (t) = 
	& -\Omega^2 X^{\text{cl}} (t) - g \sqrt{2 \Omega} ( \langle n (t) \rangle_{cl} - 1) - \Omega \Gamma (t) V (t) \nonumber
	\\
	& + \sqrt{\Omega} \xi (t) \;.
\end{align}
Here we have replaced the generic electronic operator $\langle \bar O (t) \rangle_{cl} = \langle n (t) \rangle_{cl} - 1$ in Eq.~\eqref{force}. The damping $\Gamma (t) = -2 g^2 \partial_\omega \text{Im} \chi^R_{cl} (t,\omega) |_{\omega=0}$ and the second moment of the white noise $\xi (t)$ ($\langle \xi (t) \rangle=0$, $\langle \xi (t) \xi (t') \rangle = K (t) \delta (t ,t')$, with $K (t)=-g^2\text{Im} \chi_{cl}^K (t,\omega)|_{\omega=0}$) can be computed from $\chi^R_{cl} (t,\omega)$ and $\chi^K_{cl} (t,\omega)$, which correspond to the density-density correlation functions $\chi_{cl}(t,t')=-i\langle T_C \hat n(t) \hat n(t')\rangle^\text{con}$. In the Anderson Holstein model, the action with the classical phonon field is quadratic in the electronic variables, so that the density-density correlation functions  can be factorized in terms of the electronic Green's function $G_{cl} (t, t')$ in the presence of the fluctuating field $X^{\text{cl}} (t)$. One obtains
\begin{align}
	& \chi^R_{cl} (t, t') = - i [ G^R_{cl} (t, t') G^K_{cl} (t', t) + G^K_{cl} (t, t') G^A_{cl} (t', t) ] \;,
	\\
	& \chi^K_{cl} (t, t') = - i [ G^K_{cl} (t, t') G^K_{cl} (t', t) + G^R_{cl} (t, t') G^A_{cl} (t', t) \nonumber 
	\\
	& \ \ \ \ \ \ \ \  \ \ \ \ \ \ \ \ \ + G^A_{cl} (t, t') G^R_{cl} (t', t) ] \;,
\end{align}
from which $\chi^R_{cl} (t,\omega)$ and $\chi^K_{cl} (t,\omega)$ are obtained by a Wigner transform. In the simulations, the Green's function is obtained from the QBE, see Sec.~\eqref{sec:QBE}.

In order to extract observables like $P(X)$ in the semiclassical stochastic approach, we simulate different trajectories with different noise realizations $\xi (t)$. For details about the selection of the initial configuration $X^\text{cl} (t=0)$ for each trajectory and the subsequent time-evolution to the steady state, we refer the reader to the App.~\ref{sec:init}.
Once the system has reached the steady state, the frequency of the occurrence of the realization of $X^\text{cl}$ is measured, which is distributed according to $P (X)$. We obtain good enough statistics by considering $256$ impurities, each evolved for at least $90000$ time steps (time step $\delta_t = 0.05D$). To further reduce the statistical error, in the following we plot the symmetrized distribution $(P (X ) + P (-X ))/2$.

\section{\label{sec:Results} Results and discussion} 

For different values of the inverse temperature $\beta$ and parameters $g$, $\Omega$ in the Holstein model \eqref{hamilt_imp}, the distributions $P(X)$ of the equilibrium phonon displacements generated by the different approaches will be benchmarked against the numerically exact solution from QMC. In particular, we will compare the phonon distribution function $P(X)$ and the electronic spectra.

\subsection{Distribution functions $P(X)$ and crossover to the polaronic state}

Let us first discuss the overall physical behavior that emerges from the model. When all other parameters in the impurity Hamiltonian are fixed, the distribution $P(X)$, which is symmetric with respect to $X=0$, develops a bimodal shape as the value of the electron-phonon interaction $g$ is increased. This double-peak structure indicates the formation of a polaronic state, which can already be understood on a qualitative level from a semiclassical frozen-phonon picture (adiabatic approximation): In this picture, the adiabatic potential $V_{\text{ad}}(X)$ for the phonon is determined as the sum of the bare phonon potential $E_{\text{ph}}=\tfrac12\Omega^2 X^2$, and  the energy $E_{\text{el}}(X)$ of the electronic system with phonon displacement frozen at $X$ (see Appendix~\ref{sec:frozen_phonon} for details). The phonon distribution is then approximated by the Boltzmann distribution $P_{\text{ad}}(X) \propto \exp (-\beta V_{\text{ad}}(X))$, neglecting the kinetic energy of the phonon. A large frozen phonon displacement $X$ creates a deep impurity level for the electrons, so that the electronic ground state behaves like $E_{\text{el}}(X)\sim \text{const.} - \sqrt{2\Omega} g |X| $ for large $X$  [c.f.~Eq.~\eqref{hamilt_imp}]. (Note that the potential is symmetric with respect to $X$, because the Hamiltonian is invariant under a combined particle-hole transformation and  inversion $X\rightarrow -X$.) For large $g$, the adiabatic potential $V_{\text{ad}}(X)= E_{\text{el}}(X)+ \Omega^2 X^2/2$ therefore assumes a double-well form, which implies the bimodal distribution $P(X)$. The full solution can be understood as a polaron undergoing both quantum tunneling and thermal excitations across the potential barrier, which leads to a renormalization of the distribution $P(X)$ with respect to the adiabatic one.

\begin{figure}
	\centerline{\includegraphics[width=0.5\textwidth]{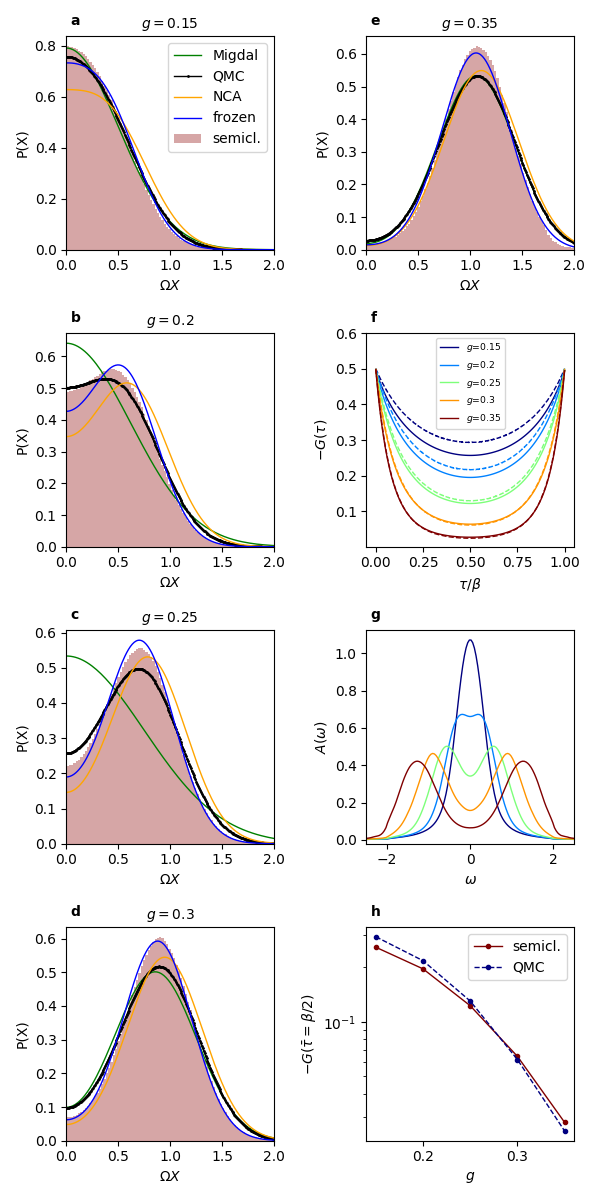}}
	\caption{(a)-(e) Equilibrium distributions $P(X)$ of the phonon displacement $X$, for increasing values of the electron-phonon interaction $g$, at inverse temperature $\beta=10$ and phonon frequency $\Omega=0.2$.  In each figure, the $P(X)$ from QMC (dotted black lines), the semiclassical stochastic approach (histograms in red), the self-consistent Migdal approximation (solid green lines), the NCA (solid orange lines), and frozen phonon approximation (solid blue lines) are shown. In the Migdal case, we show the symmetrized distribution. (f) Imaginary time Green's function $G(\tau)$ from QMC (solid lines) and the semiclassical approximation (dashed lines); (g) Equilibrium spectra from the semiclassical approximation; (h)  $G(\tau= \beta/2)$ from QMC (solid lines) and the semiclassical approximation (dashed lines). }
	\label{fig:equi}
\end{figure}

In Fig.~\ref{fig:equi}, $\Omega$ and $\beta$ are fixed, and $g$ is varied between the different panels. In the exact QMC results (error bars are of the order of the line width), we see
that as $g$ increases, the weight of the phonon distribution at $X=0$ decreases, making evident the formation of the polaronic state. The phonon distribution is described qualitatively well by the frozen phonon approach (blue lines), which supports the qualitative picture of the polaron undergoing quantum tunneling and thermal excitations between the minima of the adiabatic double well potential.  In the semiclassical approximation (histograms in red shaded color), we observe the same qualitative behavior, with some quantitative differences: The distribution generally shows a lower variance with respect to the QMC result, and reduced weight around $X=0$ in the polaronic regime. This is consistent with the interpretation that the polarons can be thermally excited across the potential barrier, like in QMC, but do not exhibit quantum tunneling between the wells at low temperatures. For the same reason, for high values of $g$, when the double-well potential is well formed (Fig.~\ref{fig:equi}d,e), the semiclassical distribution gets increasingly localized with respect to the exact QMC results. In the crossover region to the polaronic regime (Fig.~\ref{fig:equi}b), however, one can see that the semiclassical approach provides a clear improvement over the adiabatic approximation, in particular in the barrier region at $X=0$. Indeed, the frozen-phonon approximation overestimates the stability of the polaronic state since it completely neglects the thermal fluctuations of the lattice parameter $X$, which, in contrast, are included within the stochastic semiclassical theory. One can envision that, in specific parameter regimes, the difference between the two methods might become qualitative rather than merely quantitative.

By comparing the  Migdal solution with QMC, we notice that the former approaches reproduces the exact solution by construction for small values of $g$ (Fig.~\ref{fig:equi}a) while it strongly deviates from QMC at intermediate values (Fig.~\ref{fig:equi}b,c). At large values of $g$, the self-consistent Migdal result undergoes a symmetry breaking transition, where $\langle X\rangle$ acquires a nonzero expectation value; thus $P(X)$ would have a single Gaussian peak centered at $\langle \hat X \rangle$, instead of a bimodal distribution.  While a symmetry breaking is clearly not expected to happen for a single impurity model, it is interesting to note that for large $g$, the symmetrized distribution $\bar P(X) \equiv \frac{P(X)+P(-X)}{2}$, which is plotted in (Fig.~\ref{fig:equi}d,e), is close to the exact result. This qualitative behavior is expected: the Migdal theory can be viewed as a Gaussian theory that takes into account small quantum fluctuations around a mean-field solution $\langle \hat X \rangle$. With this, the symmetrized Migdal result can properly reproduce the QMC results when the exact distribution approaches a Gaussian (centered around $X=0$), or when the polaronic state is so well formed that the distribution can be approximated by two Gaussians centered around $\pm \bar X$, with $\bar X$ being the maximum of $P(X)$. Hence, it works well for small values of $g$ or when the double well potential is deep enough that tunneling is negligible and the polaron is well localized around positive or negative values of $ \bar X$. 

Finally, we consider the result obtained from the NCA approximation. These results take into account both the quantum and thermal  fluctuations of the phonon. However, due to the strong-coupling perturbative nature of the method, the system shows a strong tendency to the insulating state, leading to a significant underestimation of the critical $g$ for the onset of the polaronic regime. This result is not surprising since we are working in a parameter regime (no electron-electron repulsion) where NCA is expected to fail in reproducing the quantitative features of the transition.

\subsection{Electronic spectral functions}

In addition to the phonon variables, we also compare electronic properties obtained with QMC and the semiclassical approach. (In the semiclassical approach, electronic correlation functions are averaged over the phonon trajectories.) In Fig.~\ref{fig:equi}f we compare the Matsubara component $G(\tau)$ of the electronic Green's function, which can be calculated with QMC. This quantity is related to the real frequency spectrum by
\begin{align}
	G(\tau)=-\int \dd \omega e^{-\omega \tau} A(\omega) [1-f_\beta (\omega)],
\end{align}
so that the quantity $G(\tau =\beta/2)$ provides a good measure for the spectral weight at $\omega=0$ and low temperature.  One can thus conclude from the figure that the spectral weight at $\omega=0$ is suppressed as the polaronic regime is approached (see also Fig.~\ref{fig:equi}h). The semiclassical method provides an overall reasonable agreement with the QMC results for $G(\tau)$, which improves as  $g$ increases (see Fig.~\ref{fig:equi}f,h). Consistent with the distribution functions $P(X)$, which show a stronger polaronic localization for the semiclassical approximation, the semiclassical approximation also underestimates the electronic spectrum at $\omega=0$. 

With the semiclassical approach, one can also obtain real frequency spectra, which would have to be extracted by analytical continuation from the QMC data. As shown in Fig.~\ref{fig:equi}g, One finds an opening of the electronic gap concomitant with the appearance of the polaronic state in the phonon distribution.

\begin{figure}
	\centerline{\includegraphics[width=0.5\textwidth]{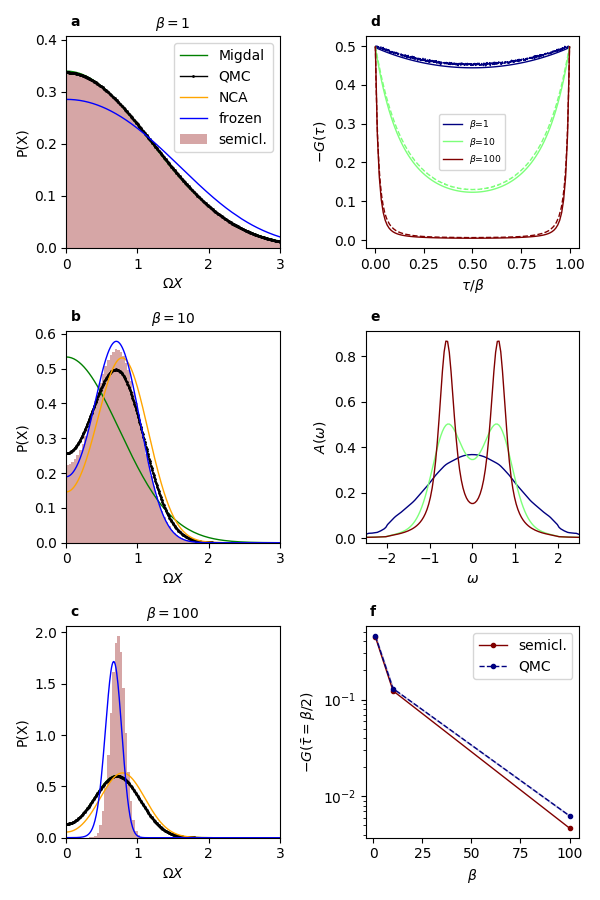}}
	\caption{(a)-(c) Equilibrium distributions $P(X)$ of the phonon displacement $X$, for increasing values of the inverse temperature $\beta$, at electron-phonon coupling $g=0.25$ and phonon frequency $\Omega=0.2$. The symbols for QMC, the semiclassical stochastic approach, self-consistent Migdal, and NCA are the same as in Fig.~\ref{fig:equi}. (d) Green's function $G(\tau)$ from QMC (solid lines) and the semiclassical approximation (dashed lines); (e) Equilibrium spectra from the semiclassical approximation; (f) $G(\tau= \beta/2)$ from QMC (solid lines) and the semiclassical approximation (dashed lines).}
	\label{fig:equi2}
\end{figure}

\subsection{Evolution with temperature}

In Fig.~\ref{fig:equi2}, we analyze the behavior of the system with decreasing temperature $T=1/\beta$, for fixed parameters  $g=0.25$  and $\Omega=0.2$. First of all, the QMC results show that the polaronic regime is reached as temperature is lowered. For the given parameters $\Omega$ and $g$ the system is in the polaronic regime (the adiabatic potential has a double well form in the ground state), but thermal excitations across the wells delocalize the polaron at high temperature. This high-temperature behavior is perfectly matched by the semiclassical result (Fig.~\ref{fig:equi2}a). As the temperature decreases, however, the semiclassical results deviate from QMC: in particular at $\beta=100$, when the thermal fluctuations are almost suppressed (Fig.~\ref{fig:equi2}c), the fact that the semiclassical approach neglects quantum tunneling makes the phonons much more localized in the two wells of the potential. Interestingly, despite that, the electronic Green's function for $\beta=100$ in the semiclassical approximation are close to the exact ones (Fig.~\ref{fig:equi2}d), suggesting that the electronic degree of freedom is mostly influenced by the position of the peak in the distribution function rather than by the phonon fluctuations in this regime. Again, the Migdal solution approaches the exact QMC results when the double well is not present (Fig.~\ref{fig:equi2}a) and when it is deep enough (after restoring the symmetry from the artificially symmetry broken state, Fig.~\ref{fig:equi2}c), while it fails to describe the bimodal distribution at intermediate temperatures (Fig.~\ref{fig:equi2}b). The NCA approximation always overestimates the position of the peak of the distribution, while for the frozen-phonon approximation the same comments as above apply  regarding the suppression of the distribution at $X=0$ and the enhancement of the value of $P(\bar X)$. By looking at the electronic properties, we notice in Fig.~\ref{fig:equi2}d that the Matsubara Green's function in the semiclassical approximation is very close to the QMC one in all the regimes and this is also confirmed in Fig.~\ref{fig:equi2}f. (The noise in the QMC data at high temperatures is due to an inefficient measurement procedure in the case of low diagram orders.) The electronic spectrum in the semiclassical approximation in Fig.~\ref{fig:equi2}e shows again the opening of the gap with the formation of the polaronic states.

\begin{figure}
	\centerline{\includegraphics[width=0.5\textwidth]{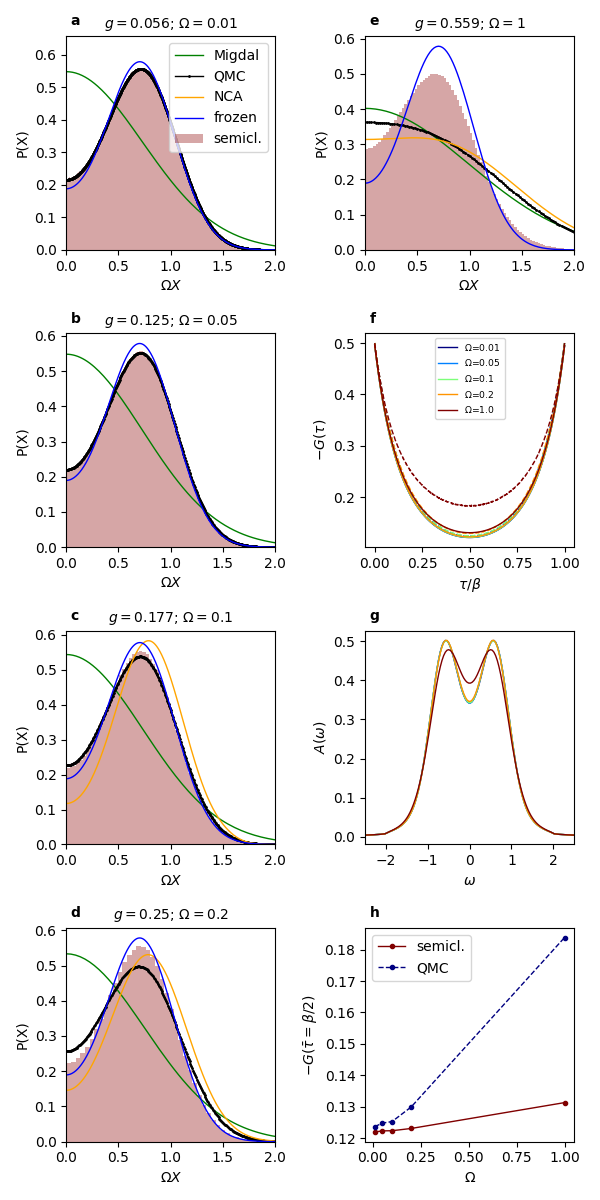}}
	\caption{(a)-(e) Equilibrium distributions $P(X)$ for different phonon frequencies $\Omega$ at a fixed value of  $g^2/\Omega=0.3125$ and at inverse temperature $\beta=10$. The symbols for QMC, the semiclassical stochastic approach, self-consistent Migdal, and NCA are the same as in Fig.~\ref{fig:equi}. For the smaller values of $\Omega$, $\Omega=0.01, 0.05$, the NCA results are not shown, because they would need a prohibitively large cutoff in the phonon Hilbert space to reach converged results with $\sqrt{\Omega} X \gg 1$. (f) Imaginary-time Green's function $G(\tau)$ from QMC (solid lines) and the semiclassical approximation (dashed lines); (g) Equilibrium spectra from the semiclassical approximation; (h) $G(\tau= \beta/2)$ from QMC (solid lines) and the semiclassical approximation (dashed lines).} 
	\label{fig:equi3}
\end{figure}

\subsection{Evolution towards the adiabatic limit}

In Fig.~\ref{fig:equi3}, we vary the phonon frequency $\Omega$ to approach the adiabatic limit ($\Omega \ll 1$), fixing $\beta$ and the ratio $g^2/\Omega$. The equilibrium position of the phonon is proportional to $g/\Omega$, so it is shifted towards lower values with increasing $\Omega$, from Fig.~\ref{fig:equi3}a to \ref{fig:equi3}e. Moreover, by increasing $\Omega$, the phonon period $2 \pi /\Omega$ decreases, while the depth of the potential well remains unchanged; in this way, the white noise approximation is less justified even when quantum tunneling is less relevant, and the semiclassical results deviate more strongly with respect to QMC (Fig.~\ref{fig:equi3}c-e). It would be interesting to see whether this deviation  can be reduced by the non-Markovian stochastic approach outlined in Sec.~\ref{nonmark}, but  implementing this is left for future research. The deviation is visible also in the electronic properties shown in Fig.~\ref{fig:equi3}f and Fig.~\ref{fig:equi3}h. 
As expected, the adiabatic approximation (blue lines) becomes more accurate for small $\Omega$, but the stochastic semiclassical theory always shows a better agreement with the QMC results, even in the regime were the adiabatic limit fails (Fig.~\ref{fig:equi3}e). This reflects the larger overestimation of the polaronic domain in the frozen phonon approximation, compared to the stochastic semiclassical theory. For the value of $g^2/\Omega$ used in this figure, the Migdal approximation fails to reproduce the formation of the polaronic state, while NCA generally overestimates the peak position of the phonon distribution. 

\section{\label{sec:Conclusion}Conclusion}

In conclusion, we have developed a microscopic formalism to treat the nonequilibrium dynamics of linearly coupled electron-phonon systems. The linear coupling is of the form $g \hat X \hat O$, where $\hat O$ is a generic electronic operator. Within this framework, the effect of the electronic fluctuations on the phonon is kept beyond the Ehrenfest dynamics, leading to a stochastic phonon evolution with damping and noise terms that are self-consistently obtained from the connected $O O$ autocorrelation functions of the electrons. The electronic dynamics in the presence of this fluctuating semiclassical distortion field is computed by means of a non-perturbative Quantum Boltzmann equation that might even be combined with a steady state nonequilibrium DMFT solver. This leads to a consistent description in which the electronic spectra and distribution functions are influenced by the stochastic dynamics of the lattice distortions, while the lattice distortions are in turn affected by the dynamics of the electrons and by their fluctuations.

To benchmark this approach, we solved the equilibrium Anderson-Holstein model with several methods (frozen phonon approximation, second order Migdal expansion and non crossing approximation) besides the stochastic semiclassical theory and numerically exact quantum Monte Carlo. We found good agreement for the phonon distribution functions  between the quantum Monte Carlo results and the stochastic semiclassical theory at higher temperatures, in particular in the crossover region to the polaronic regime. The latter would correspond to the most interesting regime close to the charge ordering temperature in a DMFT solution of the Holstein model. At low temperature, the semiclassical results overestimate the localization of the polaron due to the lack of quantum tunneling. Such quantum tunneling contributions are incorporated into multiconfigurational Ehrenfest methods~\cite{tenBrink2022}, which are however challenging for a macroscopic number of electronic degrees of freedom. A possible way of including quantum tunneling in the stochastic semiclassical theory, which would be interesting to explore in the future, is the combination with the Wentzel-Kramers-Brillouin (WKB) method, which naturally admits these processes.

The semiclassical approach can be used to address the coupled dynamics of the electrons and the lattice during photo-induced phase transitions up to long times \cite{Picano2021_CDW}, and goes beyond the phenomenological Ginzburg-Landau description. This general formulation can be extended to study more general electron-phonon couplings, such as nonlinear couplings or Jahn Teller phonons in multi-orbital models, or even to the Hubbard model, where semiclassical charge and spin fields can be introduced through Hubbard-Stratonovich transformations \cite{Dutta2022}. In the latter case, the white noise limit considered in this work might fail, and one would have to describe the dynamics of the fields by using the non-Markovian equations of motion outlined in this work.

\begin{acknowledgments}
F.G. acknowledges useful discussions with Dante M. Kennes on related topics.  M.E., F.G. and A.P.  acknowledge financial support from the ERC starting grant No.~716648. P.W. acknowledges support from ERC consolidator Grant No.~724103. The numerical calculations have been performed at the RRZE of the University Erlangen-Nuremberg.
\end{acknowledgments}

\appendix


\section{ \label{sec:frozen_phonon} Equilibrium displacement in the frozen-phonon limit}
We are interested in the equilibrium value of the displacement $X$ that minimizes the total energy of  a single impurity~\eqref{hamilt_imp}, coupled with the bath $\Delta$ in Eq.~\eqref{spectrum_Delta}, in the adiabatic limit of frozen phonons. 
The total energy is given by four contributions: the free phonon energy, the electronic kinetic energy, the electron-phonon interaction energy and, eventually, the on-site energy.
The phonon energy, i.e., the expectation value of the third term in Eq.~\eqref{hamilt_imp} reads
\begin{align}
	\label{phonon_energy}
	E_{\text{ph}}=\frac{1}{2} [\langle  \Omega^2 X^2 \rangle + \langle  P^2 \rangle ] \approx \frac{\Omega^2}{2} X^2.
\end{align}
In the frozen phonon case, the momentum of the phonon is zero since the phonon is fixed at position $X$, so $E_{\text{ph}}=\frac{\Omega^2}{2}X^2$, where $X$ is now a parameter.
Using the Keldysh formalism, the kinetic energy can be expressed as
\begin{align}
	\label{kin_energy}
	&E_{\text{kin}}=-2i [\Delta * G + G *\Delta]^<(t,t) \nonumber \\ 
	& = \frac{1}{2\pi} \int d \omega \{-2i[\Delta(\omega)G(\omega)+G(\omega)\Delta(\omega)]^< \},
\end{align}  
where $G$, the impurity's Green's function, has to be determined by solving the Dyson equation of the impurity coupled to the bath $\Delta$. 
The retarded component of $G$ thus reads 
\begin{align}
	\label{dyson_G_with_Delta}
	G^R(\omega)=[\omega+i\eta+\mu -h_{\text{loc}}-\Delta^R(\omega)]^{-1},
\end{align}
where  $h_{\text{loc}}=\sqrt{2 \Omega} g X$  is taken from Eq.~\eqref{h_loc} in the frozen-phonon approximation. 
$\eta$ is a positive small constant, $\eta \to 0^+$, that will be neglected in the following (if we want to explicitly consider it, it is enough to apply the substitution $\Im{\Delta^R(\omega)} \to \Im{\Delta^R(\omega)} - \eta $ in the equations below).
By explicitly writing the real and imaginary part of $G^R(\omega)$, we get
\begin{align}
	\label{dyson_G_with_Delta_ReIm}
	G^R(\omega)=& \frac{\omega - \bar{\omega}}{[\omega - \bar{\omega}]^2+ ( \Im  \{ \Delta^R(\omega)\} )^2} \nonumber \\
	&+i\frac{\Im \{\Delta^R(\omega)\}}{[\omega - \bar{\omega}]^2+ ( \Im \{ \Delta^R(\omega)\} )^2} ,
\end{align}
where $\bar{\omega}=-\mu+h_{\text{loc}}+ \Re \{ \Delta^R(\omega) \}$.
With this, the spectrum is obtained as $\mathcal{A}(\omega)=-\frac{1}{\pi}\Im \{G^R(\omega) \}$. We assume that the system is in equilibrium at the temperature of the bath, so the lesser component of the Green's function, $G^<$ is given by the fluctuation-dissipation theorem
\begin{align}
	\label{G_les}
	&G^<(\omega)= 2 \pi i \mathcal{A(\omega)} f_\beta (\omega),
\end{align}
with $f_\beta$ denoting the Fermi distribution function at inverse temperature $\beta$. Having determined the full $G$, we can explicitly calculate the kinetic energy by using Eq.~\eqref{kin_energy},
\begin{align}
	\label{kin_energy1}
	&E_{\text{kin}} 
	=\frac{1}{2\pi} \int d \omega \{-2i[\Delta^R(\omega)G^<(\omega) \nonumber \\
	&\hspace{11mm}+\Delta^<(\omega) G^A(\omega)+G^R(\omega) \Delta^<(\omega)+G^<(\omega) \Delta^A(\omega)] \} \nonumber \\
	&= 4 \int d \omega [ \Re \{\Delta^R(\omega)\} \mathcal{A}(\omega) + \Re\{ G^R(\omega)\} \mathcal{A}_\Delta(\omega)] f_\beta (\omega).
\end{align} 
Here $\Re \{ G^R \}$ is taken from Eq.~\eqref{dyson_G_with_Delta_ReIm}, $\mathcal{A}_\Delta$ from Eq.~\eqref{spectrum_Delta}, and 
\begin{align}
	\Re \Delta^R(\omega) =  \frac{2V^2}{D^2} [\omega - \sgn(\omega) \theta(\omega ^2-D^2)\sqrt{\omega^2-D^2}]	
	\nonumber
\end{align}
is obtained from the spectrum $\mathcal{A}_\Delta$ by a Kramers Kronig transform.

We are left at this point with just the electron-phonon interaction term, that is the expectation of the second term in Eq.~\eqref{hamilt_imp}. Within the frozen phonon approach, this gives
\begin{align}
	\label{interaction_energy}
	&E_{\text{int}}(t)= \sqrt{2 \Omega} g  X \Big [ 2  \int  d \omega \mathcal{A}(\omega)f_\beta(\omega) -1 \Big ],
\end{align}   
where the integral gives the expectation value of the electronic impurity occupation. As usual, $n=n_{\downarrow}+n_{\uparrow}$ and the factor $2$ in the last line of Eq.~\eqref{interaction_energy}, comes from the fact that we are considering the spin-symmetric phase and omitting the spin indices $G_\sigma$. Finally, the expectation value of the first term of Eq.~\eqref{hamilt_imp} reads
\begin{align}
	\label{onsite_energy}
	E_{\text{on-site}}= - \mu \langle n \rangle = - 2 \mu \int d \omega \mathcal{A}(\omega)f_\beta(\omega). 
\end{align}
The total energy
\begin{align}
	 V_{\text{ad}}(X) \equiv E_{\text{tot}}(X)=E_{\text{ph}}+E_{\text{kin}}+E_{\text{int}}+E_{\text{on-site}}
	\label{Etot_frozen}
\end{align}
defines the adiabatic potential $V_{\text{ad}}(X)$ introduced in the main text. The phonon distribution $P(X)$ is given by the Boltzmann distribution
\begin{align}
	P(X)=\frac{1}{\mathcal{Z}}e^{-\beta E_{\text{tot}}(X)},
\end{align}
where $\mathcal{Z}=\int_{-\infty}^{+\infty} \dd X e^{-\beta E_{\text{tot}}(X)}$.

\section{ \label{sec:ED} Two-site model} 
\begin{figure}
	\centerline{\includegraphics[width=0.5\textwidth]{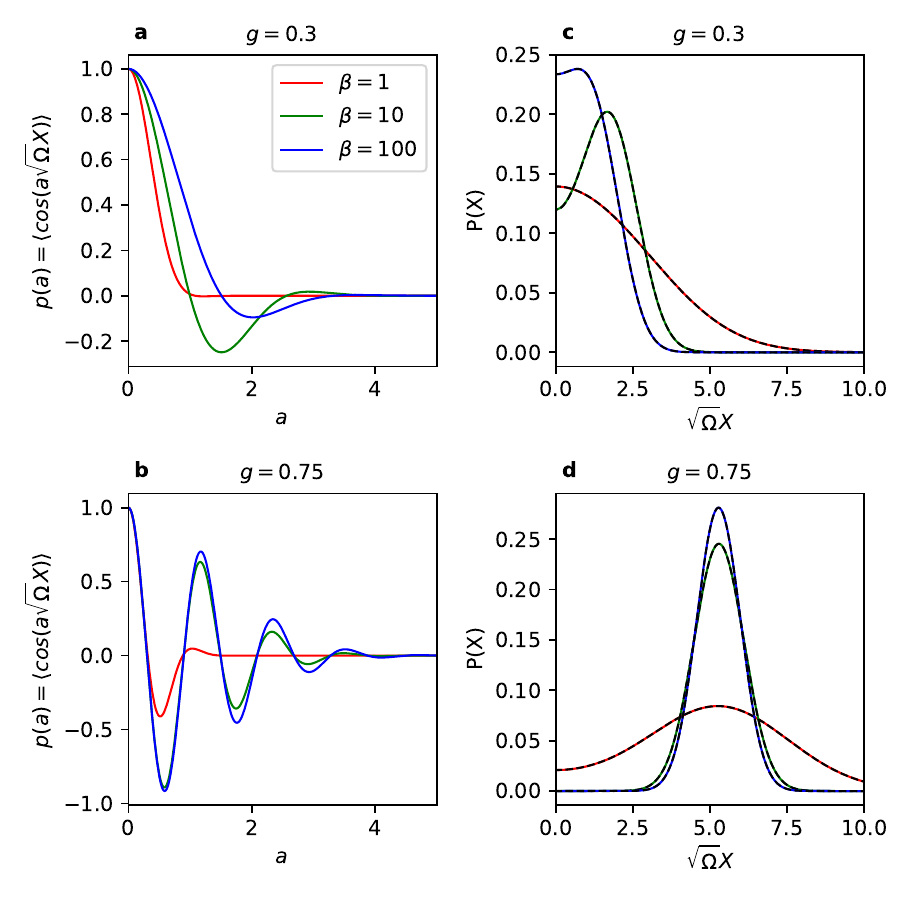}}
	\caption{a)-b): Quantum Monte Carlo distributions 
		$p(a)=\langle \text{cos} (a \sqrt{\Omega} X)\rangle_\text{MC}$ 
		for $\Omega=0.2$ 
		and $\beta=1,10,100$. Panel (a) is for $g=0.3$ and panel (b) for $g=0.75$.
		(c)-(d):
		Comparison between the distributions $P(X)$ obtained from QMC (colored solid lines) and ED (dashed black lines) for the same parameters as in (a)-(b).
	}
	\label{fig:test}
\end{figure}

Here we consider a simple impurity model where the bath is replaced by a single site, in oder to benchmark the measurements of the distribution function $P(X)$ in QMC against exact diagonalization. The Hamiltonian reads
\begin{align} 
	\label{hamilt_2imp}
	&\hat H_{\text{imp}}=
	 t' \sum_{\sigma}(f_{1,\sigma}^\dagger f_{2,\sigma} + f_{2,\sigma}^\dagger f_{1,\sigma}  )
	-\sum_{i=1,2}\sum_{\sigma} \mu_i \hat n_{i,\sigma}
	\nonumber\\
	&
	+ 
	\sqrt{2 \Omega} g \hat X ( \hat n_{1,\uparrow}+ \hat n_{1,\downarrow} -1) 
	 + \frac{1}{2}( \Omega^2 \hat X^2+ \hat P^2),
	\end{align}
where site 1 and 2 refer to the impurity and bath site, respectively.  The last term of Eq.~\eqref{hamilt_2imp} can be written as  $H_{\text{ph}}=\Omega d^\dagger d$ by recalling the definition of $\hat X=(d^\dagger+d)/\sqrt{2 \Omega}$ and  $\hat P= i \sqrt{\frac{\Omega}{2}} (d^\dagger-d)$. The parameters are chosen consistent with the ones in the main text: $\Omega=0.2$, $\mu_1=\mu_2=0$, $t'=-0.3162$, $\beta=10$, and $g=0.15,0.2,0.25,0.3,0.35$.
We solve this toy problem by exact diagonalization (ED), representing fermionic operators in the Fock space with  four possible electronic states per site ($\vert i, 0 \rangle$, $\vert i, \uparrow \rangle$, $\vert i, \downarrow \rangle$ and $\vert i, \uparrow \downarrow \rangle$, with i=1,2), and truncating the bosonic Hilbert space to a maximum number $N_{\text{ph}}$ (which is chosen large enough to obtain converged results). We calculate the probability $P(X)dX$ of finding a phonon between $X$ and $X+dX$:
\begin{align} 
	\label{P_x_ED}
	P(X)=
	&\Tr [\hat{\rho} (\mathbb{I}_{\text{el}} \otimes  \ket{X} \bra{X})] \nonumber \\
	&= \sum_{n,m}\Big(\sum_{\alpha} \bra{n,\alpha} \hat{\rho} \ket{\alpha, m}  \Big)\bra{m}\ket{X} \bra{X}\ket{n} \nonumber \\ 
	&\equiv \sum_{n,m} [\Tr_{\text{el}} \hat{\rho}]_{n,m} \bra{m}\ket{X} \bra{X}\ket{n}
	,
\end{align}
similarly to Eq.~\eqref{ph_distrib_NCA}, where $\hat \rho = \frac{1}{\mathcal{Z}} e^{-\beta \hat H_{\text{imp}}}$ with $\mathcal{Z}=\Tr[e^{-\beta \hat H_{\text{imp}}}]$ the density matrix obtained from exact diagonalization. In Eq.~\eqref{P_x_ED}, $\ket{\alpha, m} \equiv \ket{\alpha} \otimes  \ket{m} $, where $\ket{\alpha}$ is one of the 16 electronic states, while $\ket{m}$ and  $\ket{n}$ represent one of the $N_{\text{ph}}$ phononic states. $\bra{X}\ket{n} \equiv \bra{n}\ket{X}$ are the eigenfunctions of the free harmonic oscillator, defined by means of the Hermite polynomials of order $n$,
\begin{align} 
	\label{hermite_pol}
	\bra{X}\ket{n} = \Big(\frac{1}{\sqrt{\pi} 2^n n!} \Big)^{\frac{1}{2}} e^{-\frac{\Omega X^2}{2}} H_n(\sqrt{\Omega} X),
\end{align}
with $H_0(x)=1$, $H_1(x)=2x$, and $H_{n+1}(x)=2xH_n(x)-2nH_{n-1}(x)$.

 \label{sec:QMC_ED}

To illustrate and test the QMC procedure in Sec.~\ref{sec:MonteCarlo}, we show some benchmark results for the half-filled Holstein impurity coupled to a single bath site described in Eq.~\eqref{hamilt_2imp}. The hopping to the bath is chosen such that the hybridization function in imaginary time is $\Delta(\tau)=-0.05$ ($0\le \tau\le \beta$). In Fig.~\ref{fig:test}a-b, we plot the function 
$p(a)$ [Eq.~\eqref{paa}]  for $\beta=1$, $10$, $100$ and two values of $g$. This function is the output of the QMC simulation. With the post-processing procedure (\ref{postprocess}), the data in Fig.~\ref{fig:test}a-b yield the QMC phonon distributions shown in Fig.~\ref{fig:test}c-d. For this single-bath Holstein impurity model, accurate reference data for $P(X)$ with an error of less than $10^{-6}$ can be obtained using ED with $130$ phonon states. In Fig.~\ref{fig:test}c-d the dashed black lines represent these essentially exact reference data. The QMC results for $P(X)$ reproduce the ED data. Since for the QMC method, the single-bath problem is not simpler than an impurity problem with a generic hybridization function, this demonstrates the correctness and accuracy of the QMC measurement.

\section{ \label{sec:init} Equilibration procedure for the solution of the semiclassical stochastic equations} 
Here we describe the solution at the initial time $t=0$ of the stochastic equations of the type Eq.~\eqref{force}, and their successive evolution towards equilibrium.
At zero temperature, the equation of motion
\begin{align} \label{dyn_eq}
	\ddot{X}^{\text{cl}}(t) =& - \Omega^2 X^{\text{cl}}(t) - g \sqrt{2 \Omega} (\langle n (t) \rangle_{\text{cl}} - 1) \nonumber \\
 	&- \Omega \Gamma (t) \dot{X}^{\text{cl}}(t) + \sqrt{\Omega} \xi (t)
\end{align}
leads to the solution
\begin{align} \label{eq_cond}
	X^{\text{cl}} = - \frac{g \sqrt{2}}{\Omega^{\frac{3}{2}}} (\langle n \rangle_{\text{cl}} - 1) .
\end{align}
In order to determine the starting configuration of the system in Eq.~\eqref{hamilt_imp}, we initialize
all the trajectories at the same values of $X^{\text{cl}}$ and $\langle n \rangle_{\text{cl}}$ (the precise values of $X^{\text{cl}}$ and $\langle n \rangle_{\text{cl}}$ are determined by performing a self-consistent equilibrium DMFT calculation, which includes the self-consistency of the classical distortion through Eq.~\eqref{eq_cond}, besides that on the electrons). Different trajectories $i$ correspond to different realizations of the noise $\xi_i(t)$; since at $t=0$ we consider the zero-temperature (and thus zero-noise) solution, we can choose all the $X^{\text{cl}}_i(t=0)$ to be equal.

Once we have determined the starting configurations $X^{\text{cl}}_i(t=0)$, we evolve them in time through Eq.~\eqref{dyn_eq}, now at some nonzero temperature, i.e., we switch on the noise that, at each time $t>0$, may have in principle a different realization $\xi_i(t)$ for each trajectory $i$. Thanks to the noise contribution $\xi_i (t)$, each impurity $i$ experiences a different dynamics, and we evolve all of them until we observe a saturation of the standard deviation $\sigma_{X^\text{cl}} (t) = \langle X^{\text{cl},2} (t) \rangle - \langle X^{\text{cl}} (t) \rangle^2$, where the average has a statistical meaning and is performed over the $256$ impurities $i$ (or trajectories) considered in our simulations.

The state reached at this point is the equilibrium one that we consider as the starting point $\bar t$ in order to compute the relevant physical properties of the system, such as the phonon distribution function, $P(X)$. $P(X)$ is indeed an average over all the the time steps starting from the equilibrium time ($t>\bar t$) and a statistical average over all the trajectories $i$. Moreover, due to the inversion symmetry, $X \to -X$, of the system in Eq.~\eqref{hamilt_imp}, we symmetrize the $P(X)$ just calculated, by imposing $\bar P(X) \equiv \frac{P(X)+P(-X)}{2}$ (indeed, the phononic distribution must be even in the phonon variable).

\section{Non-Markovian stochastic equation}
\label{nonmark}

Although in the main text we provided a numerical solution of the stochastic equation \eqref{5637hhh} just in the white-noise limit, it is illustrating to transform the equation to a form which emphasizes the causal structure that would allow for a time-stepping solution. The starting point consists in rewriting Eq.~\eqref{eom} in the form of the two first order equations \eqref{seom1}-\eqref{seom2}. To clarify the causal structure of these equations, assume we know $V_j$ for $j<n$  and $X^{\text{cl}}_j$ for $j\le n$. The knowledge of $X$ allows to calculate  $(A^{-1})_{j,j'}$ and $\langle \bar O_{j}\rangle_{cl}$ for all $j,j'\le n$, because the latter are expectation values for the electronic problem in the presence of a time-dependent field \eqref{fluct}, which do not depend on the field at future times. It remains to be seen how this information can be used to obtain a probability distribution for $\xi_n$. Equation~\eqref{seom1} then determines the velocity $V_n$, and thus  $X^{\text{cl}}_{n+1}$ at the next time step via Eq.~\eqref{seom2}.

To construct the probability distribution of the noise, we use the fact the the matrix $A^{-1}$ is symmetric and  positive definite and can  therefore be factorized in Cholesky form,
\begin{align}
	A^{-1}=LDU,
\end{align}
where $U$ is upper triangular with $1$ on the diagonal, $D$ is diagonal and positive, and $L=U^T$ is lower triangular.
In matrix form (writing a matrix of dimension $3$ for simplicity of notation)
\begin{align}
	&A^{-1}\equiv\begin{pmatrix}
		\bar A_{00} & \bar A_{01} & \bar A_{02}  
		\\
		\bar A_{10} & \bar A_{11} & \bar A_{12} 
		\\
		\bar A_{20} & \bar A_{21} & \bar A_{22} 
	\end{pmatrix}
	=
	\nonumber
	\\
	&=\begin{pmatrix}
		1 & 0 & 0
		\\
		L_{10} & 1 & 0
		\\
		L_{20} & L_{21} & 1 
	\end{pmatrix}
	\begin{pmatrix}
		D_0& 0 & 0
		\\
		0& D_1 & 0
		\\
		0 & 0 & D_{2} 
	\end{pmatrix}\begin{pmatrix}
		1 & U_{01} & U_{02} 
		\\
		0  & 1 & U_{12} 
		\\
		0 & 0 & 1
	\end{pmatrix}.
	\label{choles}
\end{align}
Since the column index of $U$ (row index of $L$) corresponds to the time, let us denote the columns of $U$ and rows of $L$ as time slices of the respective matrices. From Eq.~\eqref{choles}, it is apparent that the $n$th time slice of $U$ and $L$, and the diagonal entry $D_{n}$ can be determined from the knowledge of all previous time slices $m<n$ and the upper $n\times n$ block of $A^{-1}$, i.e, the correlation function \eqref{p2} for times $\le n$. Moreover, with the Cholesky factorization, the inverse matrix is given by
\begin{align}
	A = U^{-1}D^{-1} L^{-1},
\end{align}
where  $L^{-1}$ is also lower triangular: Using the ansatz
\begin{align}
	\begin{pmatrix}
		1 & 0& 0
		\\
		0 & 1& 0
		\\
		0 & 0& 1
	\end{pmatrix}
	\stackrel{!}{=}
	\begin{pmatrix}
		1 & 0 & 0
		\\
		\bar L_{10} & 1 & 0
		\\
		\bar L_{20} & \bar L_{21} & 1 
	\end{pmatrix}
	\begin{pmatrix}
		1 & 0 & 0
		\\
		L_{10} &1 & 0
		\\
		L_{20} & L_{21} &1
	\end{pmatrix},
\end{align}
with the notation $(L^{-1})_{ij}\equiv \bar L_{ij}$, it is clear that the rows of $\bar L$  can be constructed successively from the time slices of $L$.
The quadratic form in the action  \eqref{5637hhh} then becomes
\begin{align}
	\sum_{j,l}
	&\xi_j A_{j,l}\xi_l 
	= 
	\sum_{p} \Big(\sum_{j\le p  }\bar L_{p,j} \xi_j\Big) D^{-1}_{p} \Big(\sum_{l\le p}\bar L_{p,l}\xi_l\Big),
\end{align}
and the $\xi$-dependent integral in  Eq.~\eqref{5637hhh} can be rewritten as
\begin{align}
	&\frac{1}{Z_\xi}\int \mathcal{D}[\xi] e^{-\frac{1}{2}\sum_{j,l} \xi_j A_{j,l}\xi_l }=
	\nonumber\\
	=&\,\, 
	\int \mathcal{D}[\xi]
	\prod_p \int d\eta_p   \frac{e^{-\frac{1}{2}D_p^{-1}\eta_p^2}}{\sqrt{2\pi D_p}} 
	\delta\Big(\eta_p - \sum_{j\le p  }\bar L_{p,j} \xi_j\Big).
	\label{kk2h23kw}
\end{align}
Here we also used that 
\begin{align}
	Z_\xi = \sqrt{(2\pi)^N\det(A)^{-1}}=\prod_p \sqrt{2\pi D_p}.
\end{align}
The delta function in the second line of Eq.~\eqref{kk2h23kw} defines a fixed relation between the variables $\eta$ and $\xi$, which can be inverted to obtain
\begin{align}
	\label{xifrometa}
	\xi_j = \sum_{p\le j  }L_{j,p} \eta_p.
\end{align}
These equations finally provide the basis for a possible time-stepping solution of the path integral \eqref{5637hhh}: Assuming that the problem has been solved up to time step $n$, i.e., $X^{\text{cl}}_j$ is known for $j\le n$, while $V_j$, $\eta_j$, $\langle \bar O_j\rangle_{cl}$, $D_j$, and the matrices $A^{-1}_{j,j'}$ and $L_{j,j'}$ are known for $j,j'<n$, the solution at the next step $n+1$ is updated as follows:
\begin{itemize}
	\item[(1)]
	Propagate the electronic problem in the presence of $X^{\text{cl}}_j$ by one time step to calculate $\langle \bar O_n\rangle_{cl}$ at time step $n$, and to extend the correlation matrix $A^{-1}_{j,l} = ig^2(\chi^K_{cl})_{j,l}$ [Eq.~\eqref{po3}] to the upper $n\times n$ block. For reasons of causality, the solution depends only on the known $X^{\text{cl}}_j$ for $j\le n$.
	\item[(2)]
	From the upper $n\times n$ block of $A^{-1}$  calculate the next time slice of the Cholesky decomposition, i.e., the $n$th row of $L$, and the diagonal weight $D_{n}$.
	\item[(3)]
	Draw a Gaussian random variable $\eta_n$ with zero mean and variance $D_n$.
	\item[(4)]
	Calculate $\xi_n$ from Eq.~\eqref{xifrometa}, using the noise $\eta$ at all previous time steps  and the $n$th time slice of $L$.
	\item[(5)]
	Update $V_{n}$ and $X^{\text{cl}}_{n+1}$ from Eqs.~\eqref{seom1} and \eqref{seom2}.
\end{itemize}


\begin{thebibliography}{46}%
	\makeatletter
	\providecommand \@ifxundefined [1]{%
		\@ifx{#1\undefined}
	}%
	\providecommand \@ifnum [1]{%
		\ifnum #1\expandafter \@firstoftwo
		\else \expandafter \@secondoftwo
		\fi
	}%
	\providecommand \@ifx [1]{%
		\ifx #1\expandafter \@firstoftwo
		\else \expandafter \@secondoftwo
		\fi
	}%
	\providecommand \natexlab [1]{#1}%
	\providecommand \enquote  [1]{``#1''}%
	\providecommand \bibnamefont  [1]{#1}%
	\providecommand \bibfnamefont [1]{#1}%
	\providecommand \citenamefont [1]{#1}%
	\providecommand \href@noop [0]{\@secondoftwo}%
	\providecommand \href [0]{\begingroup \@sanitize@url \@href}%
	\providecommand \@href[1]{\@@startlink{#1}\@@href}%
	\providecommand \@@href[1]{\endgroup#1\@@endlink}%
	\providecommand \@sanitize@url [0]{\catcode `\\12\catcode `\$12\catcode
		`\&12\catcode `\#12\catcode `\^12\catcode `\_12\catcode `\%12\relax}%
	\providecommand \@@startlink[1]{}%
	\providecommand \@@endlink[0]{}%
	\providecommand \url  [0]{\begingroup\@sanitize@url \@url }%
	\providecommand \@url [1]{\endgroup\@href {#1}{\urlprefix }}%
	\providecommand \urlprefix  [0]{URL }%
	\providecommand \Eprint [0]{\href }%
	\providecommand \doibase [0]{https://doi.org/}%
	\providecommand \selectlanguage [0]{\@gobble}%
	\providecommand \bibinfo  [0]{\@secondoftwo}%
	\providecommand \bibfield  [0]{\@secondoftwo}%
	\providecommand \translation [1]{[#1]}%
	\providecommand \BibitemOpen [0]{}%
	\providecommand \bibitemStop [0]{}%
	\providecommand \bibitemNoStop [0]{.\EOS\space}%
	\providecommand \EOS [0]{\spacefactor3000\relax}%
	\providecommand \BibitemShut  [1]{\csname bibitem#1\endcsname}%
	\let\auto@bib@innerbib\@empty
	\bibitem [{\citenamefont {Huber}\ \emph {et~al.}(2014)\citenamefont {Huber},
		\citenamefont {Mariager}, \citenamefont {Ferrer}, \citenamefont {Sch\"afer},
		\citenamefont {Johnson}, \citenamefont {Gr\"ubel}, \citenamefont {L\"ubcke},
		\citenamefont {Huber}, \citenamefont {Kubacka}, \citenamefont {Dornes},
		\citenamefont {Laulhe}, \citenamefont {Ravy}, \citenamefont {Ingold},
		\citenamefont {Beaud}, \citenamefont {Demsar},\ and\ \citenamefont
		{Johnson}}]{Huber2014}%
	\BibitemOpen
	\bibfield  {author} {\bibinfo {author} {\bibfnamefont {T.}~\bibnamefont
			{Huber}}, \bibinfo {author} {\bibfnamefont {S.~O.}\ \bibnamefont {Mariager}},
		\bibinfo {author} {\bibfnamefont {A.}~\bibnamefont {Ferrer}}, \bibinfo
		{author} {\bibfnamefont {H.}~\bibnamefont {Sch\"afer}}, \bibinfo {author}
		{\bibfnamefont {J.~A.}\ \bibnamefont {Johnson}}, \bibinfo {author}
		{\bibfnamefont {S.}~\bibnamefont {Gr\"ubel}}, \bibinfo {author}
		{\bibfnamefont {A.}~\bibnamefont {L\"ubcke}}, \bibinfo {author}
		{\bibfnamefont {L.}~\bibnamefont {Huber}}, \bibinfo {author} {\bibfnamefont
			{T.}~\bibnamefont {Kubacka}}, \bibinfo {author} {\bibfnamefont
			{C.}~\bibnamefont {Dornes}}, \bibinfo {author} {\bibfnamefont
			{C.}~\bibnamefont {Laulhe}}, \bibinfo {author} {\bibfnamefont
			{S.}~\bibnamefont {Ravy}}, \bibinfo {author} {\bibfnamefont {G.}~\bibnamefont
			{Ingold}}, \bibinfo {author} {\bibfnamefont {P.}~\bibnamefont {Beaud}},
		\bibinfo {author} {\bibfnamefont {J.}~\bibnamefont {Demsar}},\ and\ \bibinfo
		{author} {\bibfnamefont {S.~L.}\ \bibnamefont {Johnson}},\ }\bibfield
	{title} {\bibinfo {title} {{Coherent Structural Dynamics of a Prototypical
				Charge-Density-Wave-to-Metal Transition}},\ }\href
	{https://doi.org/10.1103/PhysRevLett.113.026401} {\bibfield  {journal}
		{\bibinfo  {journal} {Phys. Rev. Lett.}\ }\textbf {\bibinfo {volume} {113}},\
		\bibinfo {pages} {026401} (\bibinfo {year} {2014})}\BibitemShut {NoStop}%
	\bibitem [{\citenamefont {Perfetti}\ \emph {et~al.}(2006)\citenamefont
		{Perfetti}, \citenamefont {Loukakos}, \citenamefont {Lisowski}, \citenamefont
		{Bovensiepen}, \citenamefont {Berger}, \citenamefont {Biermann},
		\citenamefont {Cornaglia}, \citenamefont {Georges},\ and\ \citenamefont
		{Wolf}}]{Perfetti2006}%
	\BibitemOpen
	\bibfield  {author} {\bibinfo {author} {\bibfnamefont {L.}~\bibnamefont
			{Perfetti}}, \bibinfo {author} {\bibfnamefont {P.~A.}\ \bibnamefont
			{Loukakos}}, \bibinfo {author} {\bibfnamefont {M.}~\bibnamefont {Lisowski}},
		\bibinfo {author} {\bibfnamefont {U.}~\bibnamefont {Bovensiepen}}, \bibinfo
		{author} {\bibfnamefont {H.}~\bibnamefont {Berger}}, \bibinfo {author}
		{\bibfnamefont {S.}~\bibnamefont {Biermann}}, \bibinfo {author}
		{\bibfnamefont {P.~S.}\ \bibnamefont {Cornaglia}}, \bibinfo {author}
		{\bibfnamefont {A.}~\bibnamefont {Georges}},\ and\ \bibinfo {author}
		{\bibfnamefont {M.}~\bibnamefont {Wolf}},\ }\bibfield  {title} {\bibinfo
		{title} {Time evolution of the electronic structure of
			$1T\mathrm{\text{\ensuremath{-}}}{\mathrm{TaS}}_{2}$ through the
			insulator-metal transition},\ }\href
	{https://doi.org/10.1103/PhysRevLett.97.067402} {\bibfield  {journal}
		{\bibinfo  {journal} {Phys. Rev. Lett.}\ }\textbf {\bibinfo {volume} {97}},\
		\bibinfo {pages} {067402} (\bibinfo {year} {2006})}\BibitemShut {NoStop}%
	\bibitem [{\citenamefont {Zong}\ \emph {et~al.}(2018)\citenamefont {Zong},
		\citenamefont {Kogar}, \citenamefont {Bie}, \citenamefont {Rohwer},
		\citenamefont {Lee}, \citenamefont {Baldini}, \citenamefont {Erge{\c{c}}en},
		\citenamefont {Yilmaz}, \citenamefont {Freelon}, \citenamefont {Sie},
		\citenamefont {Zhou}, \citenamefont {Straquadine}, \citenamefont {Walmsley},
		\citenamefont {Dolgirev}, \citenamefont {Rozhkov}, \citenamefont {Fisher},
		\citenamefont {Jarillo-Herrero}, \citenamefont {Fine},\ and\ \citenamefont
		{Gedik}}]{Zong2018}%
	\BibitemOpen
	\bibfield  {author} {\bibinfo {author} {\bibfnamefont {A.}~\bibnamefont
			{Zong}}, \bibinfo {author} {\bibfnamefont {A.}~\bibnamefont {Kogar}},
		\bibinfo {author} {\bibfnamefont {Y.-Q.}\ \bibnamefont {Bie}}, \bibinfo
		{author} {\bibfnamefont {T.}~\bibnamefont {Rohwer}}, \bibinfo {author}
		{\bibfnamefont {C.}~\bibnamefont {Lee}}, \bibinfo {author} {\bibfnamefont
			{E.}~\bibnamefont {Baldini}}, \bibinfo {author} {\bibfnamefont
			{E.}~\bibnamefont {Erge{\c{c}}en}}, \bibinfo {author} {\bibfnamefont {M.~B.}\
			\bibnamefont {Yilmaz}}, \bibinfo {author} {\bibfnamefont {B.}~\bibnamefont
			{Freelon}}, \bibinfo {author} {\bibfnamefont {E.~J.}\ \bibnamefont {Sie}},
		\bibinfo {author} {\bibfnamefont {H.}~\bibnamefont {Zhou}}, \bibinfo {author}
		{\bibfnamefont {J.}~\bibnamefont {Straquadine}}, \bibinfo {author}
		{\bibfnamefont {P.}~\bibnamefont {Walmsley}}, \bibinfo {author}
		{\bibfnamefont {P.~E.}\ \bibnamefont {Dolgirev}}, \bibinfo {author}
		{\bibfnamefont {A.~V.}\ \bibnamefont {Rozhkov}}, \bibinfo {author}
		{\bibfnamefont {I.~R.}\ \bibnamefont {Fisher}}, \bibinfo {author}
		{\bibfnamefont {P.}~\bibnamefont {Jarillo-Herrero}}, \bibinfo {author}
		{\bibfnamefont {B.~V.}\ \bibnamefont {Fine}},\ and\ \bibinfo {author}
		{\bibfnamefont {N.}~\bibnamefont {Gedik}},\ }\bibfield  {title} {\bibinfo
		{title} {Evidence for topological defects in a photoinduced phase
			transition},\ }\href {https://doi.org/10.1038/s41567-018-0311-9} {\bibfield
		{journal} {\bibinfo  {journal} {Nature Physics}\ }\textbf {\bibinfo {volume}
			{15}},\ \bibinfo {pages} {27} (\bibinfo {year} {2018})}\BibitemShut {NoStop}%
	\bibitem [{\citenamefont {Kogar}\ \emph {et~al.}(2019)\citenamefont {Kogar},
		\citenamefont {Zong}, \citenamefont {Dolgirev}, \citenamefont {Shen},
		\citenamefont {Straquadine}, \citenamefont {Bie}, \citenamefont {Wang},
		\citenamefont {Rohwer}, \citenamefont {Tung}, \citenamefont {Yang},
		\citenamefont {Li}, \citenamefont {Yang}, \citenamefont {Weathersby},
		\citenamefont {Park}, \citenamefont {Kozina}, \citenamefont {Sie},
		\citenamefont {Wen}, \citenamefont {Jarillo-Herrero}, \citenamefont {Fisher},
		\citenamefont {Wang},\ and\ \citenamefont {Gedik}}]{Kogar2019}%
	\BibitemOpen
	\bibfield  {author} {\bibinfo {author} {\bibfnamefont {A.}~\bibnamefont
			{Kogar}}, \bibinfo {author} {\bibfnamefont {A.}~\bibnamefont {Zong}},
		\bibinfo {author} {\bibfnamefont {P.~E.}\ \bibnamefont {Dolgirev}}, \bibinfo
		{author} {\bibfnamefont {X.}~\bibnamefont {Shen}}, \bibinfo {author}
		{\bibfnamefont {J.}~\bibnamefont {Straquadine}}, \bibinfo {author}
		{\bibfnamefont {Y.-Q.}\ \bibnamefont {Bie}}, \bibinfo {author} {\bibfnamefont
			{X.}~\bibnamefont {Wang}}, \bibinfo {author} {\bibfnamefont {T.}~\bibnamefont
			{Rohwer}}, \bibinfo {author} {\bibfnamefont {I.-C.}\ \bibnamefont {Tung}},
		\bibinfo {author} {\bibfnamefont {Y.}~\bibnamefont {Yang}}, \bibinfo {author}
		{\bibfnamefont {R.}~\bibnamefont {Li}}, \bibinfo {author} {\bibfnamefont
			{J.}~\bibnamefont {Yang}}, \bibinfo {author} {\bibfnamefont {S.}~\bibnamefont
			{Weathersby}}, \bibinfo {author} {\bibfnamefont {S.}~\bibnamefont {Park}},
		\bibinfo {author} {\bibfnamefont {M.~E.}\ \bibnamefont {Kozina}}, \bibinfo
		{author} {\bibfnamefont {E.~J.}\ \bibnamefont {Sie}}, \bibinfo {author}
		{\bibfnamefont {H.}~\bibnamefont {Wen}}, \bibinfo {author} {\bibfnamefont
			{P.}~\bibnamefont {Jarillo-Herrero}}, \bibinfo {author} {\bibfnamefont
			{I.~R.}\ \bibnamefont {Fisher}}, \bibinfo {author} {\bibfnamefont
			{X.}~\bibnamefont {Wang}},\ and\ \bibinfo {author} {\bibfnamefont
			{N.}~\bibnamefont {Gedik}},\ }\bibfield  {title} {\bibinfo {title}
		{Light-induced charge density wave in {LaTe}$_3$},\ }\href
	{https://doi.org/10.1038/s41567-019-0705-3} {\bibfield  {journal} {\bibinfo
			{journal} {Nature Physics}\ }\textbf {\bibinfo {volume} {16}},\ \bibinfo
		{pages} {159} (\bibinfo {year} {2019})}\BibitemShut {NoStop}%
	\bibitem [{\citenamefont {Maklar}\ \emph {et~al.}(2021)\citenamefont {Maklar},
		\citenamefont {Windsor}, \citenamefont {Nicholson}, \citenamefont {Puppin},
		\citenamefont {Walmsley}, \citenamefont {Esposito}, \citenamefont {Porer},
		\citenamefont {Rittmann}, \citenamefont {Leuenberger}, \citenamefont {Kubli},
		\citenamefont {Savoini}, \citenamefont {Abreu}, \citenamefont {Johnson},
		\citenamefont {Beaud}, \citenamefont {Ingold}, \citenamefont {Staub},
		\citenamefont {Fisher}, \citenamefont {Ernstorfer}, \citenamefont {Wolf},\
		and\ \citenamefont {Rettig}}]{Maklar2021}%
	\BibitemOpen
	\bibfield  {author} {\bibinfo {author} {\bibfnamefont {J.}~\bibnamefont
			{Maklar}}, \bibinfo {author} {\bibfnamefont {Y.~W.}\ \bibnamefont {Windsor}},
		\bibinfo {author} {\bibfnamefont {C.~W.}\ \bibnamefont {Nicholson}}, \bibinfo
		{author} {\bibfnamefont {M.}~\bibnamefont {Puppin}}, \bibinfo {author}
		{\bibfnamefont {P.}~\bibnamefont {Walmsley}}, \bibinfo {author}
		{\bibfnamefont {V.}~\bibnamefont {Esposito}}, \bibinfo {author}
		{\bibfnamefont {M.}~\bibnamefont {Porer}}, \bibinfo {author} {\bibfnamefont
			{J.}~\bibnamefont {Rittmann}}, \bibinfo {author} {\bibfnamefont
			{D.}~\bibnamefont {Leuenberger}}, \bibinfo {author} {\bibfnamefont
			{M.}~\bibnamefont {Kubli}}, \bibinfo {author} {\bibfnamefont
			{M.}~\bibnamefont {Savoini}}, \bibinfo {author} {\bibfnamefont
			{E.}~\bibnamefont {Abreu}}, \bibinfo {author} {\bibfnamefont {S.~L.}\
			\bibnamefont {Johnson}}, \bibinfo {author} {\bibfnamefont {P.}~\bibnamefont
			{Beaud}}, \bibinfo {author} {\bibfnamefont {G.}~\bibnamefont {Ingold}},
		\bibinfo {author} {\bibfnamefont {U.}~\bibnamefont {Staub}}, \bibinfo
		{author} {\bibfnamefont {I.~R.}\ \bibnamefont {Fisher}}, \bibinfo {author}
		{\bibfnamefont {R.}~\bibnamefont {Ernstorfer}}, \bibinfo {author}
		{\bibfnamefont {M.}~\bibnamefont {Wolf}},\ and\ \bibinfo {author}
		{\bibfnamefont {L.}~\bibnamefont {Rettig}},\ }\bibfield  {title} {\bibinfo
		{title} {Nonequilibrium charge-density-wave order beyond the thermal limit},\
	}\href {https://doi.org/10.1038/s41467-021-22778-w} {\bibfield  {journal}
		{\bibinfo  {journal} {Nature Communications}\ }\textbf {\bibinfo {volume}
			{12}},\ \bibinfo {pages} {2499} (\bibinfo {year} {2021})}\BibitemShut
	{NoStop}%
	\bibitem [{\citenamefont {Wall}\ \emph {et~al.}(2018)\citenamefont {Wall},
		\citenamefont {Yang}, \citenamefont {Vidas}, \citenamefont {Chollet},
		\citenamefont {Glownia}, \citenamefont {Kozina}, \citenamefont {Katayama},
		\citenamefont {Henighan}, \citenamefont {Jiang}, \citenamefont {Miller},
		\citenamefont {Reis}, \citenamefont {Boatner}, \citenamefont {Delaire},\ and\
		\citenamefont {Trigo}}]{Wall2018}%
	\BibitemOpen
	\bibfield  {author} {\bibinfo {author} {\bibfnamefont {S.}~\bibnamefont
			{Wall}}, \bibinfo {author} {\bibfnamefont {S.}~\bibnamefont {Yang}}, \bibinfo
		{author} {\bibfnamefont {L.}~\bibnamefont {Vidas}}, \bibinfo {author}
		{\bibfnamefont {M.}~\bibnamefont {Chollet}}, \bibinfo {author} {\bibfnamefont
			{J.~M.}\ \bibnamefont {Glownia}}, \bibinfo {author} {\bibfnamefont
			{M.}~\bibnamefont {Kozina}}, \bibinfo {author} {\bibfnamefont
			{T.}~\bibnamefont {Katayama}}, \bibinfo {author} {\bibfnamefont
			{T.}~\bibnamefont {Henighan}}, \bibinfo {author} {\bibfnamefont
			{M.}~\bibnamefont {Jiang}}, \bibinfo {author} {\bibfnamefont {T.~A.}\
			\bibnamefont {Miller}}, \bibinfo {author} {\bibfnamefont {D.~A.}\
			\bibnamefont {Reis}}, \bibinfo {author} {\bibfnamefont {L.~A.}\ \bibnamefont
			{Boatner}}, \bibinfo {author} {\bibfnamefont {O.}~\bibnamefont {Delaire}},\
		and\ \bibinfo {author} {\bibfnamefont {M.}~\bibnamefont {Trigo}},\ }\bibfield
	{title} {\bibinfo {title} {{Ultrafast disordering of vanadium dimers in
				photoexcited VO$_2$}},\ }\href
	{https://doi.org/10.1126/science.aau3873} {\bibfield  {journal} {\bibinfo
			{journal} {Science}\ }\textbf {\bibinfo {volume} {362}},\ \bibinfo {pages}
		{572} (\bibinfo {year} {2018})}\BibitemShut {NoStop}%
	\bibitem [{\citenamefont {Perez-Salinas}\ \emph {et~al.}(2022)\citenamefont
		{Perez-Salinas}, \citenamefont {Johnson}, \citenamefont {Prabhakaran},\ and\
		\citenamefont {Wall}}]{PerezSalinas2022}%
	\BibitemOpen
	\bibfield  {author} {\bibinfo {author} {\bibfnamefont {D.}~\bibnamefont
			{Perez-Salinas}}, \bibinfo {author} {\bibfnamefont {A.~S.}\ \bibnamefont
			{Johnson}}, \bibinfo {author} {\bibfnamefont {D.}~\bibnamefont
			{Prabhakaran}},\ and\ \bibinfo {author} {\bibfnamefont {S.}~\bibnamefont
			{Wall}},\ }\bibfield  {title} {\bibinfo {title} {Multi-mode excitation drives
			disorder during the ultrafast melting of a C4-symmetry-broken phase},\ }\href
	{https://doi.org/10.1038/s41467-021-27819-y} {\bibfield  {journal} {\bibinfo
			{journal} {Nature Communications}\ }\textbf {\bibinfo {volume} {13}},\
		\bibinfo {pages} {238} (\bibinfo {year} {2022})}\BibitemShut {NoStop}%
	\bibitem [{\citenamefont {Beaud}\ \emph {et~al.}(2014)\citenamefont {Beaud},
		\citenamefont {Caviezel}, \citenamefont {Mariager}, \citenamefont {Rettig},
		\citenamefont {Ingold}, \citenamefont {Dornes}, \citenamefont {Huang},
		\citenamefont {Johnson}, \citenamefont {Radovic}, \citenamefont {Huber},
		\citenamefont {Kubacka}, \citenamefont {Ferrer}, \citenamefont {Lemke},
		\citenamefont {Chollet}, \citenamefont {Zhu}, \citenamefont {Glownia},
		\citenamefont {Sikorski}, \citenamefont {Robert}, \citenamefont {Wadati},
		\citenamefont {Nakamura}, \citenamefont {Kawasaki}, \citenamefont {Tokura},
		\citenamefont {Johnson},\ and\ \citenamefont {Staub}}]{Beaud2014}%
	\BibitemOpen
	\bibfield  {author} {\bibinfo {author} {\bibfnamefont {P.}~\bibnamefont
			{Beaud}}, \bibinfo {author} {\bibfnamefont {A.}~\bibnamefont {Caviezel}},
		\bibinfo {author} {\bibfnamefont {S.~O.}\ \bibnamefont {Mariager}}, \bibinfo
		{author} {\bibfnamefont {L.}~\bibnamefont {Rettig}}, \bibinfo {author}
		{\bibfnamefont {G.}~\bibnamefont {Ingold}}, \bibinfo {author} {\bibfnamefont
			{C.}~\bibnamefont {Dornes}}, \bibinfo {author} {\bibfnamefont {S.-W.}\
			\bibnamefont {Huang}}, \bibinfo {author} {\bibfnamefont {J.~A.}\ \bibnamefont
			{Johnson}}, \bibinfo {author} {\bibfnamefont {M.}~\bibnamefont {Radovic}},
		\bibinfo {author} {\bibfnamefont {T.}~\bibnamefont {Huber}}, \bibinfo
		{author} {\bibfnamefont {T.}~\bibnamefont {Kubacka}}, \bibinfo {author}
		{\bibfnamefont {A.}~\bibnamefont {Ferrer}}, \bibinfo {author} {\bibfnamefont
			{H.~T.}\ \bibnamefont {Lemke}}, \bibinfo {author} {\bibfnamefont
			{M.}~\bibnamefont {Chollet}}, \bibinfo {author} {\bibfnamefont
			{D.}~\bibnamefont {Zhu}}, \bibinfo {author} {\bibfnamefont {J.~M.}\
			\bibnamefont {Glownia}}, \bibinfo {author} {\bibfnamefont {M.}~\bibnamefont
			{Sikorski}}, \bibinfo {author} {\bibfnamefont {A.}~\bibnamefont {Robert}},
		\bibinfo {author} {\bibfnamefont {H.}~\bibnamefont {Wadati}}, \bibinfo
		{author} {\bibfnamefont {M.}~\bibnamefont {Nakamura}}, \bibinfo {author}
		{\bibfnamefont {M.}~\bibnamefont {Kawasaki}}, \bibinfo {author}
		{\bibfnamefont {Y.}~\bibnamefont {Tokura}}, \bibinfo {author} {\bibfnamefont
			{S.~L.}\ \bibnamefont {Johnson}},\ and\ \bibinfo {author} {\bibfnamefont
			{U.}~\bibnamefont {Staub}},\ }\bibfield  {title} {\bibinfo {title} {A
			time-dependent order parameter for ultrafast photoinduced phase
			transitions},\ }\href {https://doi.org/10.1038/nmat4046} {\bibfield
		{journal} {\bibinfo  {journal} {Nature Materials}\ }\textbf {\bibinfo
			{volume} {13}},\ \bibinfo {pages} {923} (\bibinfo {year} {2014})}\BibitemShut
	{NoStop}%
	\bibitem [{\citenamefont {Gerasimenko}\ \emph {et~al.}(2019)\citenamefont
		{Gerasimenko}, \citenamefont {Vaskivskyi}, \citenamefont {Litskevich},
		\citenamefont {Ravnik}, \citenamefont {Vodeb}, \citenamefont {Diego},
		\citenamefont {Kabanov},\ and\ \citenamefont {Mihailovic}}]{Gerasimenko2019}%
	\BibitemOpen
	\bibfield  {author} {\bibinfo {author} {\bibfnamefont {Y.~A.}\ \bibnamefont
			{Gerasimenko}}, \bibinfo {author} {\bibfnamefont {I.}~\bibnamefont
			{Vaskivskyi}}, \bibinfo {author} {\bibfnamefont {M.}~\bibnamefont
			{Litskevich}}, \bibinfo {author} {\bibfnamefont {J.}~\bibnamefont {Ravnik}},
		\bibinfo {author} {\bibfnamefont {J.}~\bibnamefont {Vodeb}}, \bibinfo
		{author} {\bibfnamefont {M.}~\bibnamefont {Diego}}, \bibinfo {author}
		{\bibfnamefont {V.}~\bibnamefont {Kabanov}},\ and\ \bibinfo {author}
		{\bibfnamefont {D.}~\bibnamefont {Mihailovic}},\ }\bibfield  {title}
	{\bibinfo {title} {{Quantum jamming transition to a correlated electron glass
				in 1T-TaS$_2$}},\ }\href {https://doi.org/10.1038/s41563-019-0423-3}
	{\bibfield  {journal} {\bibinfo  {journal} {Nature Materials}\ }\textbf
		{\bibinfo {volume} {18}},\ \bibinfo {pages} {1078} (\bibinfo {year}
		{2019})}\BibitemShut {NoStop}%
	\bibitem [{\citenamefont {Ichikawa}\ \emph {et~al.}(2011)\citenamefont
		{Ichikawa}, \citenamefont {Nozawa}, \citenamefont {Sato}, \citenamefont
		{Tomita}, \citenamefont {Ichiyanagi}, \citenamefont {Chollet}, \citenamefont
		{Guerin}, \citenamefont {Dean}, \citenamefont {Cavalleri}, \citenamefont
		{Adachi}, \citenamefont {Arima}, \citenamefont {Sawa}, \citenamefont
		{Ogimoto}, \citenamefont {Nakamura}, \citenamefont {Tamaki}, \citenamefont
		{Miyano},\ and\ \citenamefont {Koshihara}}]{Ichikawa2011}%
	\BibitemOpen
	\bibfield  {author} {\bibinfo {author} {\bibfnamefont {H.}~\bibnamefont
			{Ichikawa}}, \bibinfo {author} {\bibfnamefont {S.}~\bibnamefont {Nozawa}},
		\bibinfo {author} {\bibfnamefont {T.}~\bibnamefont {Sato}}, \bibinfo {author}
		{\bibfnamefont {A.}~\bibnamefont {Tomita}}, \bibinfo {author} {\bibfnamefont
			{K.}~\bibnamefont {Ichiyanagi}}, \bibinfo {author} {\bibfnamefont
			{M.}~\bibnamefont {Chollet}}, \bibinfo {author} {\bibfnamefont
			{L.}~\bibnamefont {Guerin}}, \bibinfo {author} {\bibfnamefont
			{N.}~\bibnamefont {Dean}}, \bibinfo {author} {\bibfnamefont {A.}~\bibnamefont
			{Cavalleri}}, \bibinfo {author} {\bibfnamefont {S.-i.}\ \bibnamefont
			{Adachi}}, \bibinfo {author} {\bibfnamefont {T.-h.}\ \bibnamefont {Arima}},
		\bibinfo {author} {\bibfnamefont {H.}~\bibnamefont {Sawa}}, \bibinfo {author}
		{\bibfnamefont {Y.}~\bibnamefont {Ogimoto}}, \bibinfo {author} {\bibfnamefont
			{M.}~\bibnamefont {Nakamura}}, \bibinfo {author} {\bibfnamefont
			{R.}~\bibnamefont {Tamaki}}, \bibinfo {author} {\bibfnamefont
			{K.}~\bibnamefont {Miyano}},\ and\ \bibinfo {author} {\bibfnamefont {S.-y.}\
			\bibnamefont {Koshihara}},\ }\bibfield  {title} {\bibinfo {title} {Transient
			photoinduced `hidden' phase in a manganite},\ }\href
	{https://doi.org/10.1038/nmat2929} {\bibfield  {journal} {\bibinfo  {journal}
			{Nature Materials}\ }\textbf {\bibinfo {volume} {10}},\ \bibinfo {pages}
		{101} (\bibinfo {year} {2011})}\BibitemShut {NoStop}%
	\bibitem [{\citenamefont {Stojchevska}\ \emph {et~al.}(2014)\citenamefont
		{Stojchevska}, \citenamefont {Vaskivskyi}, \citenamefont {Mertelj},
		\citenamefont {Kusar}, \citenamefont {Svetin}, \citenamefont {Brazovskii},\
		and\ \citenamefont {Mihailovic}}]{Stojchevska2014}%
	\BibitemOpen
	\bibfield  {author} {\bibinfo {author} {\bibfnamefont {L.}~\bibnamefont
			{Stojchevska}}, \bibinfo {author} {\bibfnamefont {I.}~\bibnamefont
			{Vaskivskyi}}, \bibinfo {author} {\bibfnamefont {T.}~\bibnamefont {Mertelj}},
		\bibinfo {author} {\bibfnamefont {P.}~\bibnamefont {Kusar}}, \bibinfo
		{author} {\bibfnamefont {D.}~\bibnamefont {Svetin}}, \bibinfo {author}
		{\bibfnamefont {S.}~\bibnamefont {Brazovskii}},\ and\ \bibinfo {author}
		{\bibfnamefont {D.}~\bibnamefont {Mihailovic}},\ }\bibfield  {title}
	{\bibinfo {title} {{Ultrafast Switching to a Stable Hidden Quantum State in
				an Electronic Crystal}},\ }\href {https://doi.org/10.1126/science.1241591}
	{\bibfield  {journal} {\bibinfo  {journal} {Science}\ }\textbf {\bibinfo
			{volume} {344}},\ \bibinfo {pages} {177} (\bibinfo {year}
		{2014})}\BibitemShut {NoStop}%
	\bibitem [{\citenamefont {Jeckelmann}\ \emph {et~al.}(1999)\citenamefont
		{Jeckelmann}, \citenamefont {Zhang},\ and\ \citenamefont
		{White}}]{Jeckelmann1999}%
	\BibitemOpen
	\bibfield  {author} {\bibinfo {author} {\bibfnamefont {E.}~\bibnamefont
			{Jeckelmann}}, \bibinfo {author} {\bibfnamefont {C.}~\bibnamefont {Zhang}},\
		and\ \bibinfo {author} {\bibfnamefont {S.~R.}\ \bibnamefont {White}},\
	}\bibfield  {title} {\bibinfo {title} {{Metal-insulator transition in the
				one-dimensional Holstein model at half filling}},\ }\href
	{https://doi.org/10.1103/PhysRevB.60.7950} {\bibfield  {journal} {\bibinfo
			{journal} {Phys. Rev. B}\ }\textbf {\bibinfo {volume} {60}},\ \bibinfo
		{pages} {7950} (\bibinfo {year} {1999})}\BibitemShut {NoStop}%
	\bibitem [{\citenamefont {Ejima}\ and\ \citenamefont
		{Fehske}(2009)}]{Ejima2009}%
	\BibitemOpen
	\bibfield  {author} {\bibinfo {author} {\bibfnamefont {S.}~\bibnamefont
			{Ejima}}\ and\ \bibinfo {author} {\bibfnamefont {H.}~\bibnamefont {Fehske}},\
	}\bibfield  {title} {\bibinfo {title} {{Luttinger parameters and momentum
				distribution function for the half-filled spinless fermion Holstein model: A
				{DMRG} approach}},\ }\href {https://doi.org/10.1209/0295-5075/87/27001}
	{\bibfield  {journal} {\bibinfo  {journal} {{EPL} (Europhysics Letters)}\
		}\textbf {\bibinfo {volume} {87}},\ \bibinfo {pages} {27001} (\bibinfo {year}
		{2009})}\BibitemShut {NoStop}%
	\bibitem [{\citenamefont {Jansen}\ \emph {et~al.}(2020)\citenamefont {Jansen},
		\citenamefont {Bon\ifmmode~\check{c}\else \v{c}\fi{}a},\ and\ \citenamefont
		{Heidrich-Meisner}}]{Jansen2020}%
	\BibitemOpen
	\bibfield  {author} {\bibinfo {author} {\bibfnamefont {D.}~\bibnamefont
			{Jansen}}, \bibinfo {author} {\bibfnamefont {J.}~\bibnamefont
			{Bon\ifmmode~\check{c}\else \v{c}\fi{}a}},\ and\ \bibinfo {author}
		{\bibfnamefont {F.}~\bibnamefont {Heidrich-Meisner}},\ }\bibfield  {title}
	{\bibinfo {title} {Finite-temperature density-matrix renormalization group
			method for electron-phonon systems: Thermodynamics and Holstein-polaron
			spectral functions},\ }\href {https://doi.org/10.1103/PhysRevB.102.165155}
	{\bibfield  {journal} {\bibinfo  {journal} {Phys. Rev. B}\ }\textbf {\bibinfo
			{volume} {102}},\ \bibinfo {pages} {165155} (\bibinfo {year}
		{2020})}\BibitemShut {NoStop}%
	\bibitem [{\citenamefont {Vidmar}\ \emph {et~al.}(2011)\citenamefont {Vidmar},
		\citenamefont {Bon\ifmmode~\check{c}\else \v{c}\fi{}a}, \citenamefont
		{Tohyama},\ and\ \citenamefont {Maekawa}}]{Vidmar2011}%
	\BibitemOpen
	\bibfield  {author} {\bibinfo {author} {\bibfnamefont {L.}~\bibnamefont
			{Vidmar}}, \bibinfo {author} {\bibfnamefont {J.}~\bibnamefont
			{Bon\ifmmode~\check{c}\else \v{c}\fi{}a}}, \bibinfo {author} {\bibfnamefont
			{T.}~\bibnamefont {Tohyama}},\ and\ \bibinfo {author} {\bibfnamefont
			{S.}~\bibnamefont {Maekawa}},\ }\bibfield  {title} {\bibinfo {title} {Quantum
			dynamics of a driven correlated system coupled to phonons},\ }\href
	{https://doi.org/10.1103/PhysRevLett.107.246404} {\bibfield  {journal}
		{\bibinfo  {journal} {Phys. Rev. Lett.}\ }\textbf {\bibinfo {volume} {107}},\
		\bibinfo {pages} {246404} (\bibinfo {year} {2011})}\BibitemShut {NoStop}%
	\bibitem [{\citenamefont {De~Filippis}\ \emph {et~al.}(2012)\citenamefont
		{De~Filippis}, \citenamefont {Cataudella}, \citenamefont {Nowadnick},
		\citenamefont {Devereaux}, \citenamefont {Mishchenko},\ and\ \citenamefont
		{Nagaosa}}]{DeFilippis2012}%
	\BibitemOpen
	\bibfield  {author} {\bibinfo {author} {\bibfnamefont {G.}~\bibnamefont
			{De~Filippis}}, \bibinfo {author} {\bibfnamefont {V.}~\bibnamefont
			{Cataudella}}, \bibinfo {author} {\bibfnamefont {E.~A.}\ \bibnamefont
			{Nowadnick}}, \bibinfo {author} {\bibfnamefont {T.~P.}\ \bibnamefont
			{Devereaux}}, \bibinfo {author} {\bibfnamefont {A.~S.}\ \bibnamefont
			{Mishchenko}},\ and\ \bibinfo {author} {\bibfnamefont {N.}~\bibnamefont
			{Nagaosa}},\ }\bibfield  {title} {\bibinfo {title} {Quantum dynamics of the
			Hubbard-Holstein model in equilibrium and nonequilibrium: Application to
			pump-probe phenomena},\ }\href
	{https://doi.org/10.1103/PhysRevLett.109.176402} {\bibfield  {journal}
		{\bibinfo  {journal} {Phys. Rev. Lett.}\ }\textbf {\bibinfo {volume} {109}},\
		\bibinfo {pages} {176402} (\bibinfo {year} {2012})}\BibitemShut {NoStop}%
	\bibitem [{\citenamefont {Dorfner}\ \emph {et~al.}(2015)\citenamefont
		{Dorfner}, \citenamefont {Vidmar}, \citenamefont {Brockt}, \citenamefont
		{Jeckelmann},\ and\ \citenamefont {Heidrich-Meisner}}]{Dorfner2015}%
	\BibitemOpen
	\bibfield  {author} {\bibinfo {author} {\bibfnamefont {F.}~\bibnamefont
			{Dorfner}}, \bibinfo {author} {\bibfnamefont {L.}~\bibnamefont {Vidmar}},
		\bibinfo {author} {\bibfnamefont {C.}~\bibnamefont {Brockt}}, \bibinfo
		{author} {\bibfnamefont {E.}~\bibnamefont {Jeckelmann}},\ and\ \bibinfo
		{author} {\bibfnamefont {F.}~\bibnamefont {Heidrich-Meisner}},\ }\bibfield
	{title} {\bibinfo {title} {Real-time decay of a highly excited charge carrier
			in the one-dimensional Holstein model},\ }\href
	{https://doi.org/10.1103/PhysRevB.91.104302} {\bibfield  {journal} {\bibinfo
			{journal} {Phys. Rev. B}\ }\textbf {\bibinfo {volume} {91}},\ \bibinfo
		{pages} {104302} (\bibinfo {year} {2015})}\BibitemShut {NoStop}%
	\bibitem [{\citenamefont {Bon\ifmmode~\check{c}\else \v{c}\fi{}a}\ \emph
		{et~al.}(1999)\citenamefont {Bon\ifmmode~\check{c}\else \v{c}\fi{}a},
		\citenamefont {Trugman},\ and\ \citenamefont {Batisti\ifmmode~\acute{c}\else
			\'{c}\fi{}}}]{Bonca1999}%
	\BibitemOpen
	\bibfield  {author} {\bibinfo {author} {\bibfnamefont {J.}~\bibnamefont
			{Bon\ifmmode~\check{c}\else \v{c}\fi{}a}}, \bibinfo {author} {\bibfnamefont
			{S.~A.}\ \bibnamefont {Trugman}},\ and\ \bibinfo {author} {\bibfnamefont
			{I.}~\bibnamefont {Batisti\ifmmode~\acute{c}\else \'{c}\fi{}}},\ }\bibfield
	{title} {\bibinfo {title} {Holstein polaron},\ }\href
	{https://doi.org/10.1103/PhysRevB.60.1633} {\bibfield  {journal} {\bibinfo
			{journal} {Phys. Rev. B}\ }\textbf {\bibinfo {volume} {60}},\ \bibinfo
		{pages} {1633} (\bibinfo {year} {1999})}\BibitemShut {NoStop}%
	\bibitem [{\citenamefont {Sous}\ \emph {et~al.}(2021)\citenamefont {Sous},
		\citenamefont {Kloss}, \citenamefont {Kennes}, \citenamefont {Reichman},\
		and\ \citenamefont {Millis}}]{Sous2021}%
	\BibitemOpen
	\bibfield  {author} {\bibinfo {author} {\bibfnamefont {J.}~\bibnamefont
			{Sous}}, \bibinfo {author} {\bibfnamefont {B.}~\bibnamefont {Kloss}},
		\bibinfo {author} {\bibfnamefont {D.~M.}\ \bibnamefont {Kennes}}, \bibinfo
		{author} {\bibfnamefont {D.~R.}\ \bibnamefont {Reichman}},\ and\ \bibinfo
		{author} {\bibfnamefont {A.~J.}\ \bibnamefont {Millis}},\ }\bibfield  {title}
	{\bibinfo {title} {Phonon-induced disorder in dynamics of optically pumped
			metals from nonlinear electron-phonon coupling},\ }\href
	{https://doi.org/10.1038/s41467-021-26030-3} {\bibfield  {journal} {\bibinfo
			{journal} {Nature Communications}\ }\textbf {\bibinfo {volume} {12}},\
		\bibinfo {pages} {5803} (\bibinfo {year} {2021})}\BibitemShut {NoStop}%
	\bibitem [{\citenamefont {Jansen}\ \emph {et~al.}(2021)\citenamefont {Jansen},
		\citenamefont {Jooss},\ and\ \citenamefont {Heidrich-Meisner}}]{Jansen2021}%
	\BibitemOpen
	\bibfield  {author} {\bibinfo {author} {\bibfnamefont {D.}~\bibnamefont
			{Jansen}}, \bibinfo {author} {\bibfnamefont {C.}~\bibnamefont {Jooss}},\ and\
		\bibinfo {author} {\bibfnamefont {F.}~\bibnamefont {Heidrich-Meisner}},\
	}\bibfield  {title} {\bibinfo {title} {Charge density wave breakdown in a
			heterostructure with electron-phonon coupling},\ }\href
	{https://doi.org/10.1103/PhysRevB.104.195116} {\bibfield  {journal} {\bibinfo
			{journal} {Phys. Rev. B}\ }\textbf {\bibinfo {volume} {104}},\ \bibinfo
		{pages} {195116} (\bibinfo {year} {2021})}\BibitemShut {NoStop}%
	\bibitem [{\citenamefont {Aoki}\ \emph {et~al.}(2014)\citenamefont {Aoki},
		\citenamefont {Tsuji}, \citenamefont {Eckstein}, \citenamefont {Kollar},
		\citenamefont {Oka},\ and\ \citenamefont {Werner}}]{Aoki2014}%
	\BibitemOpen
	\bibfield  {author} {\bibinfo {author} {\bibfnamefont {H.}~\bibnamefont
			{Aoki}}, \bibinfo {author} {\bibfnamefont {N.}~\bibnamefont {Tsuji}},
		\bibinfo {author} {\bibfnamefont {M.}~\bibnamefont {Eckstein}}, \bibinfo
		{author} {\bibfnamefont {M.}~\bibnamefont {Kollar}}, \bibinfo {author}
		{\bibfnamefont {T.}~\bibnamefont {Oka}},\ and\ \bibinfo {author}
		{\bibfnamefont {P.}~\bibnamefont {Werner}},\ }\bibfield  {title} {\bibinfo
		{title} {Nonequilibrium dynamical mean-field theory and its applications},\
	}\href {https://doi.org/10.1103/RevModPhys.86.779} {\bibfield  {journal}
		{\bibinfo  {journal} {Rev. Mod. Phys.}\ }\textbf {\bibinfo {volume} {86}},\
		\bibinfo {pages} {779} (\bibinfo {year} {2014})}\BibitemShut {NoStop}%
	\bibitem [{\citenamefont {Assaad}\ and\ \citenamefont
		{Lang}(2007)}]{Assaad2007}%
	\BibitemOpen
	\bibfield  {author} {\bibinfo {author} {\bibfnamefont {F.~F.}\ \bibnamefont
			{Assaad}}\ and\ \bibinfo {author} {\bibfnamefont {T.~C.}\ \bibnamefont
			{Lang}},\ }\bibfield  {title} {\bibinfo {title} {{Diagrammatic determinantal
				quantum Monte Carlo methods: Projective schemes and applications to the
				Hubbard-Holstein model}},\ }\href
	{https://doi.org/10.1103/PhysRevB.76.035116} {\bibfield  {journal} {\bibinfo
			{journal} {Phys. Rev. B}\ }\textbf {\bibinfo {volume} {76}},\ \bibinfo
		{pages} {035116} (\bibinfo {year} {2007})}\BibitemShut {NoStop}%
	\bibitem [{\citenamefont {Werner}\ and\ \citenamefont
		{Millis}(2007)}]{Werner2007}%
	\BibitemOpen
	\bibfield  {author} {\bibinfo {author} {\bibfnamefont {P.}~\bibnamefont
			{Werner}}\ and\ \bibinfo {author} {\bibfnamefont {A.~J.}\ \bibnamefont
			{Millis}},\ }\bibfield  {title} {\bibinfo {title} {{Efficient Dynamical Mean
				Field Simulation of the Holstein-Hubbard Model}},\ }\href
	{https://doi.org/10.1103/PhysRevLett.99.146404} {\bibfield  {journal}
		{\bibinfo  {journal} {Phys. Rev. Lett.}\ }\textbf {\bibinfo {volume} {99}},\
		\bibinfo {pages} {146404} (\bibinfo {year} {2007})}\BibitemShut {NoStop}%
	\bibitem [{\citenamefont {Murakami}\ \emph {et~al.}(2015)\citenamefont
		{Murakami}, \citenamefont {Werner}, \citenamefont {Tsuji},\ and\
		\citenamefont {Aoki}}]{Murakami2015}%
	\BibitemOpen
	\bibfield  {author} {\bibinfo {author} {\bibfnamefont {Y.}~\bibnamefont
			{Murakami}}, \bibinfo {author} {\bibfnamefont {P.}~\bibnamefont {Werner}},
		\bibinfo {author} {\bibfnamefont {N.}~\bibnamefont {Tsuji}},\ and\ \bibinfo
		{author} {\bibfnamefont {H.}~\bibnamefont {Aoki}},\ }\bibfield  {title}
	{\bibinfo {title} {{Interaction quench in the Holstein model: Thermalization
				crossover from electron- to phonon-dominated relaxation}},\ }\href
	{https://doi.org/10.1103/PhysRevB.91.045128} {\bibfield  {journal} {\bibinfo
			{journal} {Phys. Rev. B}\ }\textbf {\bibinfo {volume} {91}},\ \bibinfo
		{pages} {045128} (\bibinfo {year} {2015})}\BibitemShut {NoStop}%
	\bibitem [{\citenamefont {Randi}\ \emph {et~al.}(2017)\citenamefont {Randi},
		\citenamefont {Esposito}, \citenamefont {Giusti}, \citenamefont {Misochko},
		\citenamefont {Parmigiani}, \citenamefont {Fausti},\ and\ \citenamefont
		{Eckstein}}]{Randi2017}%
	\BibitemOpen
	\bibfield  {author} {\bibinfo {author} {\bibfnamefont {F.}~\bibnamefont
			{Randi}}, \bibinfo {author} {\bibfnamefont {M.}~\bibnamefont {Esposito}},
		\bibinfo {author} {\bibfnamefont {F.}~\bibnamefont {Giusti}}, \bibinfo
		{author} {\bibfnamefont {O.}~\bibnamefont {Misochko}}, \bibinfo {author}
		{\bibfnamefont {F.}~\bibnamefont {Parmigiani}}, \bibinfo {author}
		{\bibfnamefont {D.}~\bibnamefont {Fausti}},\ and\ \bibinfo {author}
		{\bibfnamefont {M.}~\bibnamefont {Eckstein}},\ }\bibfield  {title} {\bibinfo
		{title} {{Probing the Fluctuations of Optical Properties in Time-Resolved
				Spectroscopy}},\ }\href {https://doi.org/10.1103/PhysRevLett.119.187403}
	{\bibfield  {journal} {\bibinfo  {journal} {Phys. Rev. Lett.}\ }\textbf
		{\bibinfo {volume} {119}},\ \bibinfo {pages} {187403} (\bibinfo {year}
		{2017})}\BibitemShut {NoStop}%
	\bibitem [{\citenamefont {Eckstein}\ and\ \citenamefont
		{Werner}(2010)}]{Eckstein2010}%
	\BibitemOpen
	\bibfield  {author} {\bibinfo {author} {\bibfnamefont {M.}~\bibnamefont
			{Eckstein}}\ and\ \bibinfo {author} {\bibfnamefont {P.}~\bibnamefont
			{Werner}},\ }\bibfield  {title} {\bibinfo {title} {Nonequilibrium dynamical
			mean-field calculations based on the noncrossing approximation and its
			generalizations},\ }\href {https://doi.org/10.1103/PhysRevB.82.115115}
	{\bibfield  {journal} {\bibinfo  {journal} {Phys. Rev. B}\ }\textbf {\bibinfo
			{volume} {82}},\ \bibinfo {pages} {115115} (\bibinfo {year}
		{2010})}\BibitemShut {NoStop}%
	\bibitem [{\citenamefont {Werner}\ and\ \citenamefont
		{Eckstein}(2013)}]{Werner2013}%
	\BibitemOpen
	\bibfield  {author} {\bibinfo {author} {\bibfnamefont {P.}~\bibnamefont
			{Werner}}\ and\ \bibinfo {author} {\bibfnamefont {M.}~\bibnamefont
			{Eckstein}},\ }\bibfield  {title} {\bibinfo {title} {{Phonon-enhanced
				relaxation and excitation in the Holstein-Hubbard model}},\ }\href
	{https://doi.org/10.1103/PhysRevB.88.165108} {\bibfield  {journal} {\bibinfo
			{journal} {Phys. Rev. B}\ }\textbf {\bibinfo {volume} {88}},\ \bibinfo
		{pages} {165108} (\bibinfo {year} {2013})}\BibitemShut {NoStop}%
	\bibitem [{\citenamefont {Werner}\ and\ \citenamefont
		{Eckstein}(2015)}]{Werner2015}%
	\BibitemOpen
	\bibfield  {author} {\bibinfo {author} {\bibfnamefont {P.}~\bibnamefont
			{Werner}}\ and\ \bibinfo {author} {\bibfnamefont {M.}~\bibnamefont
			{Eckstein}},\ }\bibfield  {title} {\bibinfo {title} {Field-induced polaron
			formation in the Holstein-Hubbard model},\ }\href
	{https://doi.org/10.1209/0295-5075/109/37002} {\bibfield  {journal} {\bibinfo
			{journal} {{EPL} (Europhysics Letters)}\ }\textbf {\bibinfo {volume}
			{109}},\ \bibinfo {pages} {37002} (\bibinfo {year} {2015})}\BibitemShut
	{NoStop}%
	\bibitem [{\citenamefont {Chen}\ \emph {et~al.}(2016)\citenamefont {Chen},
		\citenamefont {Cohen}, \citenamefont {Millis},\ and\ \citenamefont
		{Reichman}}]{Chen2016}%
	\BibitemOpen
	\bibfield  {author} {\bibinfo {author} {\bibfnamefont {H.-T.}\ \bibnamefont
			{Chen}}, \bibinfo {author} {\bibfnamefont {G.}~\bibnamefont {Cohen}},
		\bibinfo {author} {\bibfnamefont {A.~J.}\ \bibnamefont {Millis}},\ and\
		\bibinfo {author} {\bibfnamefont {D.~R.}\ \bibnamefont {Reichman}},\
	}\bibfield  {title} {\bibinfo {title} {{Anderson-Holstein model in two
				flavors of the noncrossing approximation}},\ }\href
	{https://doi.org/10.1103/PhysRevB.93.174309} {\bibfield  {journal} {\bibinfo
			{journal} {Phys. Rev. B}\ }\textbf {\bibinfo {volume} {93}},\ \bibinfo
		{pages} {174309} (\bibinfo {year} {2016})}\BibitemShut {NoStop}%
	\bibitem [{\citenamefont {Grandi}\ \emph {et~al.}(2021)\citenamefont {Grandi},
		\citenamefont {Li},\ and\ \citenamefont {Eckstein}}]{Grandi2021b}%
	\BibitemOpen
	\bibfield  {author} {\bibinfo {author} {\bibfnamefont {F.}~\bibnamefont
			{Grandi}}, \bibinfo {author} {\bibfnamefont {J.}~\bibnamefont {Li}},\ and\
		\bibinfo {author} {\bibfnamefont {M.}~\bibnamefont {Eckstein}},\ }\bibfield
	{title} {\bibinfo {title} {Ultrafast Mott transition driven by nonlinear
			electron-phonon interaction},\ }\href
	{https://doi.org/10.1103/PhysRevB.103.L041110} {\bibfield  {journal}
		{\bibinfo  {journal} {Phys. Rev. B}\ }\textbf {\bibinfo {volume} {103}},\
		\bibinfo {pages} {L041110} (\bibinfo {year} {2021})}\BibitemShut {NoStop}%
	\bibitem [{\citenamefont {Kamenev}(2011)}]{KamenevBook}%
	\BibitemOpen
	\bibfield  {author} {\bibinfo {author} {\bibfnamefont {A.}~\bibnamefont
			{Kamenev}},\ }\href {https://doi.org/10.1017/CBO9781139003667} {\emph
		{\bibinfo {title} {{Field Theory of Non-Equilibrium Systems}}}}\ (\bibinfo
	{publisher} {Cambridge University Press},\ \bibinfo {year}
	{2011})\BibitemShut {NoStop}%
	\bibitem [{\citenamefont {Mitra}\ \emph {et~al.}(2005)\citenamefont {Mitra},
		\citenamefont {Aleiner},\ and\ \citenamefont {Millis}}]{Mitra2005}%
	\BibitemOpen
	\bibfield  {author} {\bibinfo {author} {\bibfnamefont {A.}~\bibnamefont
			{Mitra}}, \bibinfo {author} {\bibfnamefont {I.}~\bibnamefont {Aleiner}},\
		and\ \bibinfo {author} {\bibfnamefont {A.~J.}\ \bibnamefont {Millis}},\
	}\bibfield  {title} {\bibinfo {title} {Semiclassical analysis of the
			nonequilibrium local polaron},\ }\href
	{https://doi.org/10.1103/PhysRevLett.94.076404} {\bibfield  {journal}
		{\bibinfo  {journal} {Phys. Rev. Lett.}\ }\textbf {\bibinfo {volume} {94}},\
		\bibinfo {pages} {076404} (\bibinfo {year} {2005})}\BibitemShut {NoStop}%
	\bibitem [{\citenamefont {Picano}\ \emph
		{et~al.}(2021{\natexlab{a}})\citenamefont {Picano}, \citenamefont {Grandi},\
		and\ \citenamefont {Eckstein}}]{Picano2021_CDW}%
	\BibitemOpen
	\bibfield  {author} {\bibinfo {author} {\bibfnamefont {A.}~\bibnamefont
			{Picano}}, \bibinfo {author} {\bibfnamefont {F.}~\bibnamefont {Grandi}},\
		and\ \bibinfo {author} {\bibfnamefont {M.}~\bibnamefont {Eckstein}},\ }\href
	{https://doi.org/10.48550/ARXIV.2112.15323} {\bibinfo {title} {Inhomogeneous
			disordering at a photo-induced charge density wave transition}} (\bibinfo
	{year} {2021}{\natexlab{a}})\BibitemShut {NoStop}%
	\bibitem [{\citenamefont {Yonemitsu}\ and\ \citenamefont
		{Maeshima}(2009)}]{Yonemitsu2009}%
	\BibitemOpen
	\bibfield  {author} {\bibinfo {author} {\bibfnamefont {K.}~\bibnamefont
			{Yonemitsu}}\ and\ \bibinfo {author} {\bibfnamefont {N.}~\bibnamefont
			{Maeshima}},\ }\bibfield  {title} {\bibinfo {title} {Coupling-dependent rate
			of energy transfer from photoexcited Mott insulators to lattice vibrations},\
	}\href {https://doi.org/10.1103/PhysRevB.79.125118} {\bibfield  {journal}
		{\bibinfo  {journal} {Phys. Rev. B}\ }\textbf {\bibinfo {volume} {79}},\
		\bibinfo {pages} {125118} (\bibinfo {year} {2009})}\BibitemShut {NoStop}%
	\bibitem [{\citenamefont {Petrovic}\ \emph {et~al.}(2022)\citenamefont
		{Petrovic}, \citenamefont {Weber},\ and\ \citenamefont
		{Freericks}}]{Petrovic2022}%
	\BibitemOpen
	\bibfield  {author} {\bibinfo {author} {\bibfnamefont {M.~D.}\ \bibnamefont
			{Petrovic}}, \bibinfo {author} {\bibfnamefont {M.}~\bibnamefont {Weber}},\
		and\ \bibinfo {author} {\bibfnamefont {J.~K.}\ \bibnamefont {Freericks}},\
	}\href {https://doi.org/10.48550/ARXIV.2203.11880} {\bibinfo {title}
		{Theoretical description of time-resolved photoemission in
			charge-density-wave materials out to long times}} (\bibinfo {year}
	{2022})\BibitemShut {NoStop}%
	\bibitem [{\citenamefont {Weber}\ and\ \citenamefont
		{Freericks}(2022)}]{Weber2022}%
	\BibitemOpen
	\bibfield  {author} {\bibinfo {author} {\bibfnamefont {M.}~\bibnamefont
			{Weber}}\ and\ \bibinfo {author} {\bibfnamefont {J.~K.}\ \bibnamefont
			{Freericks}},\ }\bibfield  {title} {\bibinfo {title} {Real-time evolution of
			static electron-phonon models in time-dependent electric fields},\ }\href
	{https://doi.org/10.1103/PhysRevE.105.025301} {\bibfield  {journal} {\bibinfo
			{journal} {Phys. Rev. E}\ }\textbf {\bibinfo {volume} {105}},\ \bibinfo
		{pages} {025301} (\bibinfo {year} {2022})}\BibitemShut {NoStop}%
	\bibitem [{\citenamefont {Osterkorn}\ and\ \citenamefont
		{Kehrein}(2022)}]{Osterkorn2022}%
	\BibitemOpen
	\bibfield  {author} {\bibinfo {author} {\bibfnamefont {A.}~\bibnamefont
			{Osterkorn}}\ and\ \bibinfo {author} {\bibfnamefont {S.}~\bibnamefont
			{Kehrein}},\ }\href {https://doi.org/10.48550/ARXIV.2205.06620} {\bibinfo
		{title} {Photoinduced prethermal order parameter dynamics in the
			two-dimensional large-$n$ hubbard-heisenberg model}} (\bibinfo {year}
	{2022})\BibitemShut {NoStop}%
	\bibitem [{\citenamefont {Bakshi}\ \emph {et~al.}(2022)\citenamefont {Bakshi},
		\citenamefont {Bose}, \citenamefont {Dutta},\ and\ \citenamefont
		{Majumdar}}]{Bakshi2022}%
	\BibitemOpen
	\bibfield  {author} {\bibinfo {author} {\bibfnamefont {S.~S.}\ \bibnamefont
			{Bakshi}}, \bibinfo {author} {\bibfnamefont {D.}~\bibnamefont {Bose}},
		\bibinfo {author} {\bibfnamefont {A.}~\bibnamefont {Dutta}},\ and\ \bibinfo
		{author} {\bibfnamefont {P.}~\bibnamefont {Majumdar}},\ }\href
	{https://doi.org/10.48550/ARXIV.2205.14710} {\bibinfo {title} {Nonequilibrium
			dynamics of suppression, revival, and loss of charge order in a laser pumped
			electron-phonon system}} (\bibinfo {year} {2022})\BibitemShut {NoStop}%
	\bibitem [{\citenamefont {Picano}\ \emph
		{et~al.}(2021{\natexlab{b}})\citenamefont {Picano}, \citenamefont {Li},\ and\
		\citenamefont {Eckstein}}]{Picano2021}%
	\BibitemOpen
	\bibfield  {author} {\bibinfo {author} {\bibfnamefont {A.}~\bibnamefont
			{Picano}}, \bibinfo {author} {\bibfnamefont {J.}~\bibnamefont {Li}},\ and\
		\bibinfo {author} {\bibfnamefont {M.}~\bibnamefont {Eckstein}},\ }\bibfield
	{title} {\bibinfo {title} {{Quantum Boltzmann equation for strongly
				correlated electrons}},\ }\href {https://doi.org/10.1103/PhysRevB.104.085108}
	{\bibfield  {journal} {\bibinfo  {journal} {Phys. Rev. B}\ }\textbf {\bibinfo
			{volume} {104}},\ \bibinfo {pages} {085108} (\bibinfo {year}
		{2021}{\natexlab{b}})}\BibitemShut {NoStop}%
	\bibitem [{\citenamefont {Grandi}\ \emph {et~al.}(2020)\citenamefont {Grandi},
		\citenamefont {Amaricci},\ and\ \citenamefont {Fabrizio}}]{Grandi2019}%
	\BibitemOpen
	\bibfield  {author} {\bibinfo {author} {\bibfnamefont {F.}~\bibnamefont
			{Grandi}}, \bibinfo {author} {\bibfnamefont {A.}~\bibnamefont {Amaricci}},\
		and\ \bibinfo {author} {\bibfnamefont {M.}~\bibnamefont {Fabrizio}},\
	}\bibfield  {title} {\bibinfo {title} {{Unraveling the Mott-Peierls intrigue
				in vanadium dioxide}},\ }\href
	{https://doi.org/10.1103/PhysRevResearch.2.013298} {\bibfield  {journal}
		{\bibinfo  {journal} {Phys. Rev. Research}\ }\textbf {\bibinfo {volume}
			{2}},\ \bibinfo {pages} {013298} (\bibinfo {year} {2020})}\BibitemShut
	{NoStop}%
	\bibitem [{\citenamefont {Sieberer}\ \emph {et~al.}(2016)\citenamefont
		{Sieberer}, \citenamefont {Buchhold},\ and\ \citenamefont
		{Diehl}}]{Sieberer2016}%
	\BibitemOpen
	\bibfield  {author} {\bibinfo {author} {\bibfnamefont {L.~M.}\ \bibnamefont
			{Sieberer}}, \bibinfo {author} {\bibfnamefont {M.}~\bibnamefont {Buchhold}},\
		and\ \bibinfo {author} {\bibfnamefont {S.}~\bibnamefont {Diehl}},\ }\bibfield
	{title} {\bibinfo {title} {Keldysh field theory for driven open quantum
			systems},\ }\href {https://doi.org/10.1088/0034-4885/79/9/096001} {\bibfield
		{journal} {\bibinfo  {journal} {Reports on Progress in Physics}\ }\textbf
		{\bibinfo {volume} {79}},\ \bibinfo {pages} {096001} (\bibinfo {year}
		{2016})}\BibitemShut {NoStop}%
	\bibitem [{\citenamefont {ten Brink}\ \emph {et~al.}(2022)\citenamefont {ten
			Brink}, \citenamefont {Gr\"{a}ber}, \citenamefont {Hopjan}, \citenamefont
		{Jansen}, \citenamefont {Stolpp}, \citenamefont {Heidrich-Meisner},\ and\
		\citenamefont {Bl\"{o}chl}}]{tenBrink2022}%
	\BibitemOpen
	\bibfield  {author} {\bibinfo {author} {\bibfnamefont {M.}~\bibnamefont {ten
				Brink}}, \bibinfo {author} {\bibfnamefont {S.}~\bibnamefont {Gr\"{a}ber}},
		\bibinfo {author} {\bibfnamefont {M.}~\bibnamefont {Hopjan}}, \bibinfo
		{author} {\bibfnamefont {D.}~\bibnamefont {Jansen}}, \bibinfo {author}
		{\bibfnamefont {J.}~\bibnamefont {Stolpp}}, \bibinfo {author} {\bibfnamefont
			{F.}~\bibnamefont {Heidrich-Meisner}},\ and\ \bibinfo {author} {\bibfnamefont
			{P.~E.}\ \bibnamefont {Bl\"{o}chl}},\ }\bibfield  {title} {\bibinfo {title}
		{{Real-time non-adiabatic dynamics in the one-dimensional Holstein model:
				Trajectory-based vs exact methods}},\ }\href
	{https://doi.org/10.1063/5.0092063} {\bibfield  {journal} {\bibinfo
			{journal} {The Journal of Chemical Physics}\ }\textbf {\bibinfo {volume}
			{156}},\ \bibinfo {pages} {234109} (\bibinfo {year} {2022})}\BibitemShut
	{NoStop}%
	\bibitem [{\citenamefont {Werner}\ \emph {et~al.}(2006)\citenamefont {Werner},
		\citenamefont {Comanac}, \citenamefont {de' Medici}, \citenamefont {Troyer},\
		and\ \citenamefont {Millis}}]{Werner2006}%
	\BibitemOpen
	\bibfield  {author} {\bibinfo {author} {\bibfnamefont {P.}~\bibnamefont
			{Werner}}, \bibinfo {author} {\bibfnamefont {A.}~\bibnamefont {Comanac}},
		\bibinfo {author} {\bibfnamefont {L.}~\bibnamefont {de' Medici}}, \bibinfo
		{author} {\bibfnamefont {M.}~\bibnamefont {Troyer}},\ and\ \bibinfo {author}
		{\bibfnamefont {A.~J.}\ \bibnamefont {Millis}},\ }\bibfield  {title}
	{\bibinfo {title} {{Continuous-Time Solver for Quantum Impurity Models}},\
	}\href {https://doi.org/10.1103/PhysRevLett.97.076405} {\bibfield  {journal}
		{\bibinfo  {journal} {Phys. Rev. Lett.}\ }\textbf {\bibinfo {volume} {97}},\
		\bibinfo {pages} {076405} (\bibinfo {year} {2006})}\BibitemShut {NoStop}%
	\bibitem [{\citenamefont {Lang}\ and\ \citenamefont {Firsov}(1962)}]{Lang1962}%
	\BibitemOpen
	\bibfield  {author} {\bibinfo {author} {\bibfnamefont {I.~G.}\ \bibnamefont
			{Lang}}\ and\ \bibinfo {author} {\bibfnamefont {Y.~A.}\ \bibnamefont
			{Firsov}},\ }\href@noop {} {\bibfield  {journal} {\bibinfo  {journal} {Sov.
				Phys. JETP}\ }\textbf {\bibinfo {volume} {16}},\ \bibinfo {pages} {1301}
		(\bibinfo {year} {1962})}\BibitemShut {NoStop}%
	\bibitem [{\citenamefont {Sch\"{u}ler}\ \emph {et~al.}(2020)\citenamefont
		{Sch\"{u}ler}, \citenamefont {Gole\v{z}}, \citenamefont {Murakami},
		\citenamefont {Bittner}, \citenamefont {Herrmann}, \citenamefont {Strand},
		\citenamefont {Werner},\ and\ \citenamefont {Eckstein}}]{Schuler2020}%
	\BibitemOpen
	\bibfield  {author} {\bibinfo {author} {\bibfnamefont {M.}~\bibnamefont
			{Sch\"{u}ler}}, \bibinfo {author} {\bibfnamefont {D.}~\bibnamefont
			{Gole\v{z}}}, \bibinfo {author} {\bibfnamefont {Y.}~\bibnamefont {Murakami}},
		\bibinfo {author} {\bibfnamefont {N.}~\bibnamefont {Bittner}}, \bibinfo
		{author} {\bibfnamefont {A.}~\bibnamefont {Herrmann}}, \bibinfo {author}
		{\bibfnamefont {H.~U.}\ \bibnamefont {Strand}}, \bibinfo {author}
		{\bibfnamefont {P.}~\bibnamefont {Werner}},\ and\ \bibinfo {author}
		{\bibfnamefont {M.}~\bibnamefont {Eckstein}},\ }\bibfield  {title} {\bibinfo
		{title} {Nessi: The non-equilibrium systems simulation package},\ }\href
	{https://doi.org/https://doi.org/10.1016/j.cpc.2020.107484} {\bibfield
		{journal} {\bibinfo  {journal} {Computer Physics Communications}\ }\textbf
		{\bibinfo {volume} {257}},\ \bibinfo {pages} {107484} (\bibinfo {year}
		{2020})}\BibitemShut {NoStop}%
	\bibitem [{\citenamefont {Dutta}\ and\ \citenamefont
		{Majumdar}(2022)}]{Dutta2022}%
	\BibitemOpen
	\bibfield  {author} {\bibinfo {author} {\bibfnamefont {A.}~\bibnamefont
			{Dutta}}\ and\ \bibinfo {author} {\bibfnamefont {P.}~\bibnamefont
			{Majumdar}},\ }\bibfield  {title} {\bibinfo {title} {Nonequilibrium thermal
			state of a voltage-biased Mott insulator},\ }\href
	{https://doi.org/10.1103/PhysRevB.105.075149} {\bibfield  {journal} {\bibinfo
			{journal} {Phys. Rev. B}\ }\textbf {\bibinfo {volume} {105}},\ \bibinfo
		{pages} {075149} (\bibinfo {year} {2022})}\BibitemShut {NoStop}%
\end{thebibliography}

%

\end{document}